\documentclass[aps,prb,showpacs,superscriptaddress,amssymb,twocolumn]{revtex4}
\usepackage{mathptmx}
\usepackage{bm,amsmath}
\makeatletter
\@ifundefined{pdfoutput} 
{ 
\usepackage[dvips]{graphicx,color}
}
{ 
\ifnum\pdfoutput=0\relax
\usepackage[dvips]{graphicx,color}
\fi
\ifnum\pdfoutput=1\relax
\usepackage[pdftex]{graphicx,color}
\fi
}
\makeatother
\newcommand{\beq}{\begin{equation}}
\newcommand{\eeq}{\end{equation}}
\newcommand{\beqa}{\begin{eqnarray}}
\newcommand{\eeqa}{\end{eqnarray}}
\newcommand{\nn}{\nonumber \\}

\def \e {\mathrm{e}}

\def \la {\langle}
\def \p {{\underline{p}}}
\def \q {{\underline{q}}}
\def \ra {\rangle}
\def \s {\sigma}

\def \B {{\mathcal B}}
\def \C {{\mathbb C}}
\def \Cl {{\mathcal C}}
\def \F {{\mathbb F}}
\def \I {{\mathbb I}}
\def \M {{\mathcal M}}
\def \P {{\mathcal P}}
\def \R {{\mathbb R}}
\def \S {{\mathcal S}}
\def \Z {{\mathbb Z}}

\def \H {{\mathcal H}}

\begin{document}
\title{Implementation of Clifford gates in the Ising-anyon topological quantum
computer}
\author{Andr\'e Ahlbrecht}
\affiliation{Institut f\"ur Mathematische Physik, Technische Universit\"at Braunschweig,
Mendelssohnstr. 3, 38106 Braunschweig, Germany}
\author{Lachezar S. Georgiev}
\affiliation{Institut f\"ur Mathematische Physik, Technische Universit\"at Braunschweig,
Mendelssohnstr. 3, 38106 Braunschweig, Germany}
\affiliation{Institute for Nuclear Research and Nuclear Energy, Bulgarian Academy of Sciences,
72 Tsarigradsko Chaussee, 1784 Sofia, Bulgaria}
\author{Reinhard F. Werner}
\affiliation{Institut f\"ur Mathematische Physik, Technische Universit\"at Braunschweig,
Mendelssohnstr. 3, 38106 Braunschweig, Germany}

\date{\today}
\begin{abstract}
We give a general proof for the existence and realizability of Clifford gates in
the Ising topological quantum computer.
We show that all quantum gates that can be implemented  by braiding of Ising anyons
are Clifford gates. We find that the braiding gates for two qubits exhaust the
entire two-qubit Clifford group.
Analyzing the structure of the Clifford group for $n\geq 3$ qubits we
prove that the the image of the braid group is a non-trivial subgroup of the
Clifford group so that not all Clifford gates could be implemented by braiding
in the Ising topological quantum computation scheme.
We also point out which Clifford gates cannot in general
 be realized by braiding.
\end{abstract}


\maketitle
\clearpage
\section{Introduction}
Quantum Computation (QC) \cite{nielsen-chuang} development encountered tremendous
difficulties in storing and
manipulating quantum information in real physical systems because of the
overwhelming decoherence and noise.
Topological Quantum Computation (TQC) \cite{kitaev-TQC}  is a branch of QC in which
both information storage and processing are protected by the topological
nature of the quantum computer.
In the TQC approach the quantum information is encoded in non-local topological
degrees of freedom and is
therefore inaccessible to noise and decoherence which are mainly due to local interactions.
Moreover, the quantum gates are implemented
by non-trivial topological operations which are once again protected against decoherence.
The purpose of this topological protection of qubits
and quantum gates is to improve the quantum computation hardware to such an extent
 that quantum information processing become more feasible in real physical systems. The feasibility of this approach depends on the detailed noise structure \cite{AlickiFannesHorodecki}. In this paper we work entirely in the noiseless case, and ask about the possibility of realizing certain key gates in a specific scheme.

One of the most promising TQC schemes \cite{sarma-freedman-nayak} employs the
anticipated non-Abelian braid statistics of the lowest energy excitations, called Ising
anyons, of the fractional quantum Hall state at filling factor $\nu=5/2$, which
is believed to belong to the  universality class of the Moore-Read (MR) Pfaffian
state \cite{mr}.
Due to the topological degeneracy in the two-dimensional critical Ising model,
representing the neutral degrees of freedom in the MR state, it becomes possible to realize  $n$-qubits by 
$2n+2$ Ising anyons:
the states of $2n+2$ Ising anyons are represented by conformal field theory (CFT) correlation functions 
(more precisely, chiral CFT blocks)
which happen to belong to one of the two inequivalent spinor irreducible representations (IRs) 
\cite{nayak-wilczek,read-JMP,TQC-PRL,TQC-NPB,ultimate,equiv}
of the covering group Spin$(2n+2)$ of the rotation group SO$(2n+2)$.
There are two inequivalent spinor IRs
\cite{nayak-wilczek,read-JMP,equiv,ultimate}
of  SO$(2n+2)$, of dimension $2^{n}$, which differ by their total fermion parity \cite{equiv,ultimate}.
Despite being mathematically inequivalent they appear to be equivalent from the computational
point of view \cite{equiv}, i.e., the set of matrices that could be obtained by braiding Ising anyons in both 
representations is the same.

Clifford-group gates, that are defined as those unitary operations that preserve the Pauli
group, play a central role in quantum information theory. Although the Clifford group is not sufficient for
universal QC and its computational power cannot exceed classical computers
\cite{nielsen-chuang,GottesmanHeisenberg,AaronsonGottesman}, experimental realization of the Clifford group 
with sufficient scalability would be a cornerstone of QC. Indeed, the Clifford group is of significance for quantum 
error correction\cite{GottesmanPhD} and allows the generation of entangled states, like GHZ or Cluster states, 
the latter being a prerequisite for universal quantum computation in the measurement-based 
scheme\cite{BriegelRaussendorf}.

Using the explicit representation of the braid-group generators for the exchanges
of Ising anyons one of us has constructed \cite{TQC-PRL,TQC-NPB} the entire Clifford groups
for 1 and 2 qubits in terms of braid generators for $\B_4$ and $\B_6$, respectively.
However, this approach encountered serious difficulties for embedding some
Clifford gates in systems
with three or more qubits \cite{TQC-PRL,TQC-NPB}: the topological entanglement between distant
Ising anyons induces additional phases when exchanging anyon pairs which are in
the state $|1\ra$, i.e., when exchanging pairs sharing Majorana fermions.
Therefore, it was possible to construct only a part of the Clifford group for 3 qubits.
In this paper we shall address the question whether all Clifford-group gates could be
realized by braiding of Ising anyons in the Ising TQC scheme or not.
Because, as we shall prove below, the $n$-qubit Pauli group coincides with the monodromy
subgroup
representation for $2n+2$ Ising anyons and because in general the monodromy group is a
normal subgroup of the braid group,  it naturally follows that the Ising-model braid group
representation is a subgroup of the  Clifford group. In other words, all quantum gates that
could be implemented by braiding of Ising anyons are actually Clifford gates.
Unfortunately, it also appears that not all Clifford gates could be realized by braiding
for three or more Ising qubits and we shall try to explain why. The Clifford gates that
cannot be implemented by braiding are typically the embeddings of the two-qubit SWAP gate
into larger systems.

The rest of this paper is organized as follows:
in Sect.~\ref{sec:fusion} we describe how anyonic states of matter could
be labeled by fusion paths in Bratteli diagrams and how this could be used to determine the
dimension of the computational space. In Sect.~\ref{sec:exchange} we summarize the explicit
representation of the elementary Ising-model exchange matrices as proposed by Nayak and
Wilczek \cite{nayak-wilczek}. In Sect.~\ref{sec:symplectic}
we explain the   symplectic description of the Clifford group for $n$ Ising qubits
and estimate the order of the Clifford group in order to compare it with the order of the braid-group
representation. In Sect.~\ref{sec:braiding} we analyze the relation between the Pauli group and the monodromy subgroup of the braid group and
explain why all quantum gates that could be realized
by braiding of Ising anyons are in fact Clifford gates. In Sect.~\ref{sec:2qubits} we give the
explicit braid construction of the two-qubit SWAP gate which allows to construct the entire
two-qubit Clifford group by braiding. Some important technical details are collected in several
appendices.
\section{Fusion paths: labeling the anyonic states of matter }
\label{sec:fusion}
The anyonic states of matter (should they really exist in Nature) differ from the
ordinary fermionic and bosonic states in that we need to give additional
non-local information in order to specify the quantum state. In TQC this extra information
 is expressed in terms of topological quantum numbers that are eventually
used to encode quantum information non-locally in order to gain topological
protection.
An important ingredient of any TQC scheme is the necessary degeneracy of
ground states of the multi-anyon system in presence of trapping potentials
\cite{sarma-RMP,stern-review} (the potentials that keep our computational anyons at fixed
positions).
This is equivalent to a degeneracy of the multi-anyon states, considered as
excitations corresponding to having a number of anyons at fixed positions
in the plane, over the ground state (this time without trapping potentials) \cite{sarma-RMP}.
In the CFT language, that we will use to characterize the anyonic
states as CFT correlators, the second point of view is more appropriate and
we shall speak about the degeneracy of the CFT correlation functions corresponding to
$2n+2$ Ising anyons at fixed positions. In more detail,  the CFT correlation functions
of $2n+2$ Ising anyons  at fixed positions inside a Pfaffian droplet\cite{nayak-wilczek}
 span a Hilbert space of dimension $2^n$ and are therefore appropriate for representing $n$
Ising qubits. The non-local internal quantum numbers, for  non-Abelian anyons, that would allow us to
distinguish between the different states in this \textit{computational space} are the so called \textit{fusion channels}. 
In order to make this notion more transparent let us try to explain the fusion rules of the
Ising model: the most relevant quasiparticle excitations in the low-temperature, low-energy regime, are 
described by the chiral Ising spin field $\s$ with dimension $1/16$. If two such quasiparticles, having 
non-trivial topological properties, are fused, i.e., 
taken to the same point in the coordinate plane the composite object
 would look (from far away) like another topological excitation and this is symbolically
expressed by the fusion rule
\beq \label{fusion}
\s \times \s = \I +\psi.
\eeq
This rule means that there are two distinct channels in the fusion process and if we consider a large number of identically prepared experiments
some pairs of $\s$ would behave collectively as the vacuum
(this is the $\I$ in Eq.~(\ref{fusion})) and this is called the vacuum channel, while some others would look like the Majorana fermion
(i.e., the $\psi$ in Eq.~(\ref{fusion})) and this is called the Majorana channel. 
In other words, the combined quasiparticle can be considered as a mixed state of the vacuum and a Majorana fermion.
Localized particle-like collective excitations, such as the field $\s$ in Eq.~(\ref{fusion}),
 which have more than one available fusion channels are called non-Abelian anyons.
 There are in general superselection rules which forbid creation of coherent superpositions of anyons, such as the vacuum and the Majorana fermion in the Ising model, belonging to different superselection sectors 
 and this has to be taken into account when using non-Abelian anyons for TQC.

The important point is that the topological properties of the anyon pairs  are persistent
(under some reasonable assumptions), i.e, if two quasiparticles, which are in a state characterized by a definite fusion channel,  are pulled away the pair still possesses the properties of their corresponding fusion channel and, e.g., if they are fused again after some time,
 they will produce the same  result as that determined by the original fusion channel.

When we have many Ising anyons, which we will assume to be ordered on a line, we could represent the string of $2n+2$
Ising anyons $\s$ into $n+1$ pairs $(\s,\s)_{c}$ and characterize each pair by its
fusion channel $c$.
Then the sequential composition of the fusion channels for all pairs can be described by
a \textit{fusion path} in the corresponding fusion diagram.
The  fusion diagrams that we will use are known as Bratteli diagram \cite{sarma-RMP,bratteli,stern-review} 
and represent graphically the possible results of fusion of a single  basic non-Abelian anyon to an array of 
other non-Abelian anyons usually of the same type.
\begin{figure}[htb]
\centering
\includegraphics*[bb=20 440 410 610,width=9cm]{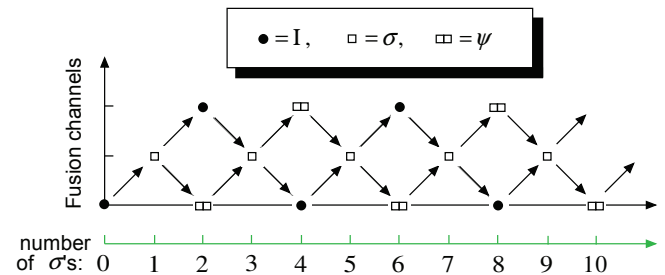}
\caption{Bratteli diagram for the Ising model, in the   $su(2)$  notation.  }
\label{fig:bratteli-2-2}
\end{figure}
One way to understand this diagram is to represent the Ising model as the affine coset\cite{CFT-book}
$\widehat{su(2)}_2/\widehat{u(1)}$ and use the  fact that the Ising model fusion rules are the same as those for 
the $\widehat{su(2)}_2$ Wess--Zumino--Witten model \cite{CFT-book}. 
Then the CFT primary fields $\s$ and  $\psi$ (of CFT dimensions $1/16$ and $1/2$ respectively),  could be labeled 
 by the reduced Young tablues for the admissible \cite{CFT-book} $\widehat{su(2)}_2$ representations
 as shown in Fig.~\ref{fig:bratteli-2-2} and define (together with the vacuum) the 3 different superselection 
sectors (anyons) of the Ising model.
 In this way the fusion rules for the Ising model can be inferred from the tensor product decomposition of the 
$su(2)$ representations. 
Each step to the right in Fig.~\ref{fig:bratteli-2-2} denotes fusing one more fundamental anyon $\s$ to the existing 
string of $\s$ anyons (whose length is determined by the $x$ coordinate)  and the arrows represent the possible 
fusion channels for this process, which are listed along the vertical axis. 
Note that the Bratteli diagram in Fig.~\ref{fig:bratteli-2-2} is finite in vertical direction which expresses the 
important property of rationality: any rational CFT, such as the critical two-dimensional Ising model, contains a 
finite number of topologically inequivalent superselection sectors (i.e., finite number of distinct anyons) which are
``closed'' under fusion. In order to build the Bratteli diagram in 
Fig.~\ref{fig:bratteli-2-2} we have to supplement Eq.~(\ref{fusion}) with the two other fusion rules 
$\psi \times \s= \s$, $\I \times\s=\s$.
All anyons, such as $\I$ and  $\psi$, from which originates only one arrow pointing to the right are Abelian, while 
those, such as $\s$, with more than one (two in this case) arrows to the right are non-Abelian.

\begin{figure}[htb]
\centering
\includegraphics*[bb=0 450 340 625,width=8.5cm]{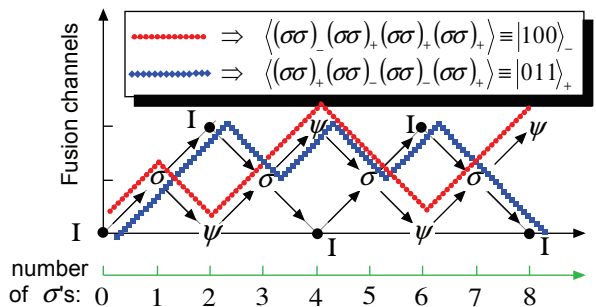}
\caption{(color online) Fusion paths in Bratteli diagram: red circles display a path
	corresponding to the
	three-qubit state $|100 \ra _-$, while the path of blue squares corresponds to the
	state $|011\ra_+$. One horizontal step to the right represents adding one more $\s$ field and
	the arrows point to the corresponding fusion channel. The black circle denotes the
	vacuum channel $\I$. The subscript $\pm$ of the computational states denotes the positive- or negative-parity representations. }
\label{fig:bratteli}
\end{figure}

In order to gain some intuition about labeling the states in the Ising model 
 we show in Fig.~\ref{fig:bratteli}  two distinct
three-qubit states  realized as fusion paths in the Bratteli diagram for 8 Ising anyons.
Notice that there are two inequivalent representations of the multi-anyon states with 8 anyons 
\cite{nayak-wilczek,ultimate,equiv}:
besides the 8 $\s$ fields inside the CFT correlation function there are also a number of Majorana fermions 
present in there (not shown explicitly in the correlation functions in Fig.~\ref{fig:bratteli} but see E
q.~(\ref{1qb-basis}) below) --
when this number  is even the state belongs to the positive-parity representation 
(in which case the 8 $\s$ fields must fuse to the identity in order for the CFT correlator to be non-zero), while 
when it is odd the state is in the negative-parity representation (and the 8 $\s$ fields must fuse to the Majorana fermion).
The Bratteli diagrams also appear to be a very
useful graphical tool for computing the dimensions of the computational spaces for TQC with anyons.
For example, the space dimension of Ising-model correlation functions with 8 anyons at fixed positions could be
read off from Fig.~\ref{fig:bratteli} as follows: each new step to the right defines a single
new fusion channel if the current number of $\s$ is even and two fusion channels if the number of $\s$s is odd.
Therefore, the number of distinct fusion paths is just $2^{\frac{\# \s}{2} -1}$,
the power of two with an exponent equal to the number of the odd $\s$-steps and finally divided by 2 because the
last step is always fixed to fuse to the vacuum or to the Majorana fermion depending on the parity of the 
representation we deal with. We see from
Fig.~\ref{fig:bratteli} that the number of independent correlation  functions of $8$ fields $\s$ is  $2^{4-1}=8$ which could be
easily generalized to the case of $2n$ $\s$ fields where the correlation functions (for fixed
positions of the anyons) span \cite{nayak-wilczek} a computational space of dimension   $2^{n-1}$.

Thus we can say that the state of a multi-anyon system can be ultimately characterized by its fusion path in the 
Bratteli diagram.
Having specified a multi-anyon state of matter as a fusion path in a Bratteli diagram
any TQC scheme requires physical processes that could initialize the multi-anyon system into a given 
$n$-qubit state and this has to be further supplemented by procedures for measuring the states of the individual qubits.
For the TQC scheme based on Ising anyons the initialization as well as the measurement
by a Fabry--Perot or Mach--Zehnder interferometer have been discussed in 
Refs.~\onlinecite{sarma-freedman-nayak,sarma-RMP,stern-review}.

Now that we have got an idea about how to encode qubits using non-Abelian anyons
we could think about executing quantum gates which are unitary transformations acting within the computational space of multi-anyon states labeled by fusion paths.
A central issue in this context is that the multi-anyon states, that we would like to use to encode qubits, are degenerate in energy (at least approximately, see Refs.~\onlinecite{sarma-RMP,stern-review}
for more detailed explanation), and separated from the rest of the excitation spectrum by a gap.
This allows us to apply  a version of the adiabatic theorem which is appropriate for the degenerate case. In simple words the adiabatic theorem in this case states that if the initial state is in the degenerate subspace that is separated by a gap from the rest of the spectrum and
we consider the adiabatic evolution, when some of the anyons are transported along  complete loops around others,
the final multi-anyon state would be again a member of the same degenerate subspace
(e.g., of ground states in presence of trapping potentials). Then the validity of the adiabatic approximation would guarantee that the transformation of the initial state into the final one is described by the action of a unitary operator which includes the Berry phase and the
explicit monodromies of the (typically multi-valued) multi-anyon states.
It is, however, possible to choose a basis of CFT blocks \cite{nayak-wilczek} in which the
Berry phases are trivial and the entire effect of the adiabatic evolution is contained
in the monodromies of the CFT correlation functions.
Thus we see that it might be possible to execute quantum gates over our topologically protected qubits by adiabatically transporting some anyons around others and these quantum operations are naturally protected against noise and decoherence.

Using the  CFT-correlators  representation we can describe the $n$-qubit Ising system
by a CFT correlation function including $2n+2$ Ising spin fields $\s$ and the
quantum-information encoding rule could be chosen to be e.g.,
\[
|c_1, \ldots, c_i, \ldots, c_n \ra \to \la (\s \s)_{c_1}
\cdots (\s \s)_{c_i} \cdots (\s \s)_{c_n}   (\s \s)_{c_0}\ra_{\mathrm{CFT}}
\]
where $c_0=c_1c_2\cdots c_n$. Notice that the $n$ qubits are encoded from left to
right starting with the first pair of $\s$ fields and there is always one extra pair,
which we shall take as the right-most one, which is inert, i.e., it contains no independent information because 
its fusion channel $c_0$ is determined by the product of the fermion parities of  the individual qubits states. In 
other words, the physical meaning of $c_i$ is the fermion parity of the pair representing the $i$-th qubit,  
the role of the inert pair is to compensate the combined fermion parity of the first $n$ pairs so that the correlation 
function is non-zero.

Since we shall use the fermion parity of the non-Abelian $\s$ fields to encode information
in the Ising qubits it is worth saying a few words about it.
The chiral spin fields $\s$ are identified with the primary fields of the Ising CFT
that intertwine between the  vacuum sector
(or the Neveu-Schwarz (NS) sector) and the so called Ramond (R) sector of the Ising model  
\cite{fst,5-2}. The conservation of the fermion parity,
combined with the fact that the Majorana fermion has a zero mode in the R-sector,
implies that Ramond sector in the Ising model is double degenerate \cite{fst,5-2}.
This means that there must exist two  Abelian
chiral spin fields $\s_+$ and $\s_-$ of CFT dimension
$1/16$ with fermion parity $+$ and $-$ respectively. However, this double degeneration of
the R-sector is incompatible with the modular invariance which is at the heart of the
fusion rules for the anyonic model. In order to guarantee the modular invariance of the
model we need to choose only one linear combination \cite{5-2} of $\s_+$ and $\s_-$ which is conventionally taken as
\beq \label{sigma}
\s=\frac{\s_+ + \s_-}{\sqrt{2}}.
\eeq
As a result of this (GSO) projection \cite{5-2} the chiral fermion parity is
spontaneously broken which is also obvious in the fusion rules (\ref{fusion}).
Note that conservation of chiral fermion parity requires that the fields $\s_\pm$
with definite parity are Abelian, i.e., their fusion rules must be
$\s_+ \times \s_+ = \I$, $\s_- \times \s_- = \I$ and $\s_+ \times \s_- = \psi$,
and the non-Abelian statistics appears only when this
symmetry is broken (which could only be spontaneous because the generator of the fermion
parity symmetry commutes with the conformal Hamiltonian).

\begin{figure}[htb]
\centering
\includegraphics*[bb=30 590 230 675,width=5cm]{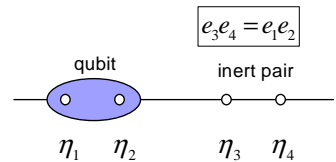}
\caption{Single qubit configuration in terms of 4 Ising quasiholes
	 corresponding  to the positive-parity representation $S_+$ of the braid group $\B_4$.}
\label{fig:1qubit}
\end{figure}

Although the only field which is believed to appear in the physical system is
(\ref{sigma}) and the chiral fermion parity is broken by the non-Abelian fusion
rule (\ref{fusion})  the fields $\s_\pm$ are still convenient for labeling our
computational basis in terms of Ising-model correlation functions. In other words,
even if $\s_\pm$ are unphysical they could efficiently label the independent functions
spanning the degenerate space of correlation functions because the product $e_1 e_2 = c=\pm 1$
of the two indices in a pair $\s_{e_1}\s_{e_2}$ determines the fusion channel $c=+1$ for the vacuum channel and $c=-1$ for the Majorana one.
Thus, the general qubit encoding scheme for Ising qubits can be represented as
\beqa\label{encoding}
|0\ra \quad & \Leftrightarrow & \quad  (\s \s)_{+1} \quad \Leftrightarrow \quad \s_+ \s_+ \quad \mathrm{and} \nn
|1\ra \quad & \Leftrightarrow & \quad  (\s \s)_{-1} \quad \Leftrightarrow  \quad \s_+ \s_-.
\eeqa
More precisely, using for instance the CFT-description of Ising qubits in the positive-parity representation,
and introducing one extra pair as illustrated in Fig.~\ref{fig:1qubit},
we could write the single-qubit computational basis as
\beqa \label{1qb-basis}
|0\ra_+ &\equiv& \la \s_+\s_+ \ \s_+\s_+  \prod_{j=1}^{2N} \psi(z_j) \ra_{\mathrm{CFT}} \nn
|1\ra_+ &\equiv& \la \s_+\s_- \ \s_+\s_-   \prod_{j=1}^{2N} \psi(z_j) \ra_{\mathrm{CFT}} ,
\eeqa
where the first pair of $\s$ fields  represents the qubit while the second one
is the non-computational inert pair of Ising anyons
compensating the fermion parity.
The total topological charge of the 4 $\s$ fields  is $+1$
which corresponds to the positive-parity spinor representation of SO$(4)$ (cf. next section)
and is denoted as a subscript of the computational basis states in
Eq.~(\ref{1qb-basis}). 
The computational basis for two Ising qubits is explicitly written in
Sect.~\ref{sec:2qubits} and for more qubits in Refs.~\onlinecite{equiv,ultimate}.
\section{Elementary exchange matrices: spinor approach}
\label{sec:exchange}
As we have already mentioned the $n$-qubit system could be represented by $2n+2$ Ising
anyons because the degeneracy of the states containing $2n+2$ anyons is \cite{nayak-wilczek}
$2^{\frac{2n+2}{2} -1}=2^n$. The most natural approach to determine the
braid matrices representing the exchanges of adjacent anyons over the (degenerate)
space of correlation functions is to use the analytic continuation of these
multi-valued analytic functions that have been used to define the qubits. 
The results of Ref.~\onlinecite{nayak-wilczek} suggested that
the multi-anyon Ising wave functions span a spinor representation of the rotation group
SO$(2n+2)$ (more precisely, one of the inequivalent spinor representations of its covering
group Spin$(2n+2)$), however, this important conjecture has not been completely proven in
Ref.~\onlinecite{nayak-wilczek}. It has later been proven rigorously in Ref.~\onlinecite{TQC-NPB} for the 
positive-parity representation of the braid group $\B_{4}$ and in Ref.~\onlinecite{ultimate} for both
positive- and negative-parity representations of the braid group $\B_{2n+2}$ representing the exchanges 
in the $n$-qubit states defined as CFT correlation functions of $2n+2$ Ising anyons.
One of the main results of  Ref.~\onlinecite{ultimate} is that the $n$-qubit representations of the
 braid-group generators $B_j^{(2n+2,\pm)}$, obtained by analytic continuation of the CFT correlation functions,
 are \textit{completely} equivalent  to those, denoted as  $R_j^{(n+1,\pm)}$, obtained from the standard Clifford-algebra realization   of the generators of the rotation group SO$(2n+2)$, i.e.,
\[
B_j^{(2n+2,\pm)}= \left( C^{(2n+2,\pm)} \right)^{-1} R_j^{(n+1,\pm)}  C^{(2n+2,\pm)}.
\]
Notice that this is not completely trivial since the matrices which establish the equivalence between
the braid generators in the wave-function representations with definite parity and those in the Clifford-algebra
realizations with the same parity are special non-singular entangling matrices (i.e., matrices which cannot be 
written as factorized tensor products of smaller ones) which are explicitly constructed in
Ref.~\onlinecite{ultimate}. For example the equivalence matrices for 3 qubits in the two representations are
$C^{(8,+)}=\textrm{diag}(-1,1,1,1,1,1,1,-1)$ while $C^{(8,-)}=i\textrm{diag}(1,1,1,-1,1,-1,-1,-1)$.
 Thus, as proven  in Ref.~\onlinecite{ultimate},  it is completely legitimate to use the Clifford-algebra 
construction of the  braid-group generators for those representing the coordinate exchanges in the CFT
 correlation functions.  The advantage of using the Clifford-algebra representation of the
 braid generators comes from the possibility to express the Pauli gates in terms of
 the squares of  elementary braid generators (cf. section \ref{sec:braiding}) 
 and then use the group-theoretic approach to find the stabilizing Clifford group.

That is why we shall review below the general Clifford-algebra construction of
the two inequivalent spinor representations of the braid group $\B_{2n}$
which will be used later to implement various quantum operations over
the Ising multi-qubit systems.
Nayak and Wilczek suggested \cite{nayak-wilczek} that the elementary
matrices representing the coordinate exchanges of $2n$ Ising quasiparticles in the
Pfaffian FQH wave functions can be expressed in terms of the gamma matrices
$\gamma_i^{(n)}$,  $1\leq i \leq 2n$, which satisfy the anticommutation relations
of the Clifford algebra
\beq \label{cr}
\left\{\gamma_i^{(n)},\gamma_j^{(n)} \right\} = 2\delta_{ij}, \quad 1\leq i,j \leq 2n.
\eeq
In more detail,
the elementary operations for the exchange of the $j$-th and $(j+1)$-th quasiparticles
could be expressed (including the correct phase) as \cite{nayak-wilczek,ivanov,equiv,ultimate}
\beq \label{R}
R_{j}^{(n)} = \e^{i\frac{\pi}{4}}
\exp\left(-\frac{\pi}{4} \gamma_j^{(n)}\gamma_{j+1}^{(n)}\right) \equiv
\frac{\e^{i\frac{\pi}{4}}}{\sqrt{2}}
\left(\I -\gamma_j^{(n)}\gamma_{j+1}^{(n)}\right).
\eeq
The $2n$  matrices $\gamma^{(n)}_j$ have dimension $2^n \times 2^n$ and
can be defined recursively in terms of the Pauli matrices
$\sigma_k$ ($k=1,2,3$) as follows  \cite{wilczek-zee}
\beqa \label{gamma}
\gamma_j^{(n+1)} &=&  \gamma_j^{(n)} \otimes \sigma_3 ,\quad 1\leq j \leq 2n \nn
\gamma_{2n+1}^{(n+1)} &=&  \I_{2^n}\otimes \sigma_1\nn
\gamma_{2n+2}^{(n+1)} &=&  \I_{2^n} \otimes\sigma_2.
\eeqa
Starting with $n=0$ as a base, where $\gamma_{1}^{(1)} =\sigma_1$ and
$\gamma_{2}^{(1)} =\sigma_2$ we could write the gamma matrices explicitly as follows
\cite{wilczek-zee,slingerland-bais}
\beqa \label{gamma-ex}
\gamma_{1}^{(n)} &=& \sigma_1 \otimes \sigma_3\otimes \cdots \otimes\sigma_3   \nn
\gamma_{2}^{(n)} &=& \sigma_2 \otimes \sigma_3\otimes \cdots \otimes\sigma_3  \nn
&\vdots& \nn
\gamma_{2j-1}^{(n)} &=& \underbrace{\I_{2} \otimes \cdots \otimes \I_{2}}_{j-1}\otimes
\ \sigma_1 \otimes \underbrace{\sigma_3\otimes\cdots \otimes\sigma_3 }_{n-j}  \nn
\gamma_{2j}^{(n)} &=&  \underbrace{\I_{2} \otimes  \cdots \otimes \I_{2}}_{j-1}\otimes
\ \sigma_2 \otimes \underbrace{\sigma_3\otimes\cdots\otimes\sigma_3}_{n-j}  \nn
&\vdots& \nn
\gamma_{2n-1}^{(n)} &=&  \I_{2^{n-1}}\otimes \sigma_1 \nn
\gamma_{2n}^{(n)} &=&  \I_{2^{n-1}} \otimes\sigma_2.
\eeqa

The ``gamma-five'' matrix $\gamma_F^{(n)}$ could be constructed by
\[
\gamma_F^{(n)} =(-i)^n \gamma_1^{(n)} \cdots  \gamma_{2n}^{(n)}=
\underbrace{\s_3\otimes \cdots \otimes \s_3}_{n}.
\]
Because $\gamma_F^{(n)}$ anticommutes with all  $\gamma_j^{(n)}$ it commutes with all
$R_{j}$ and therefore this representation of the braid group is reducible.
Since $\gamma_F^{(n)}$  has only two distinct eigenvalues, $\pm 1$, it splits into two
irreducible representations that could be obtained with the help of
the two projectors
\beq \label{P_pm}
P_{\pm}^{(n)} =\frac{\I_{2^n} \pm\gamma_F^{(n)} }{2}, \ \mathrm{where}
\ \left( P_{\pm}^{(n)}\right)^2= P_{\pm}^{(n)} =\left( P_{\pm}^{(n)}\right)^\dagger .
\eeq
In other words, the generators of the two irreducible spinor representations of the
braid group $\B_{2n}$ can be obtained by simply projecting Eq.~(\ref{R}), i.e.,
\beqa \label{R_pm}
R_{j}^{(n,\pm) } &=&\e^{i\frac{\pi}{4}}
P_{\pm}^{(n)} \exp\left(-\frac{\pi}{4} \gamma_j\gamma_{j+1}\right)P_{\pm}^{(n)}   \nn
&=& \frac{\e^{i\frac{\pi}{4}}}{\sqrt{2}}\left(\I -\gamma_j^{(n)}\gamma_{j+1}^{(n)}\right)
P_{\pm}^{(n)} .
\eeqa
It would be useful for our analysis to give a more detailed expressions for the braid generators
\beqa
R_{2j-1}^{\pm}&=&\frac{\e^{i\frac{\pi}{4}}}{\sqrt{2}}\left(\I_2 -i\s_3^{(j)}\right)
P_{\pm}^{(n)},  \quad 1\leq j \leq n\nn
R_{2j}^{\pm}&=&\frac{\e^{i\frac{\pi}{4}}}{\sqrt{2}}\left(\I_4 -i\s_2^{(j)}\otimes \s_2^{(j+1)}\right)
P_{\pm}^{(n)}\, , \, 1\leq j \leq n-1\nonumber
\eeqa
 where we have used the shorthand notation $\s_k^{(j)}$ to denote $\s_k$ acting on qubit $j$. 
A more explicit form of the matrices can be found in Ref.~\onlinecite{equiv}.
\section{Clifford group for $n$ qubits: symplectic description}
\label{sec:symplectic}
Quantum operations belonging to the Clifford group are characterized in the Heisenberg picture as the group of 
unitary operators which stabilize the Pauli group, i.e. they map an arbitrary factorized tensor product of Pauli operators 
to another factorized tensor product of Pauli operators. This property has important consequences, extending a 
suitable description of the Pauli group (cf. Eq.~(\ref{P-order})) to the Clifford group equips us with an efficient way 
of writing down such unitaries (cf. Eq.~(\ref{isomorphism})) as opposed to general unitaries where the number 
of parameters typically grows exponentially with the number of qubits. Although, this eliminates exponential 
speed-up, the Clifford group is of great relevance for quantum computation. On the one hand we can study 
basic properties of quantum mechanical systems, like entanglement generation, explicitly and on the other hand there 
are ingredients of quantum computation, e.g. error correction\cite{GottesmanPhD}, which are based on the Clifford 
group. Detailed descriptions of Clifford group quantum computations are given in 
Ref. \onlinecite{nielsen-chuang,GottesmanHeisenberg,AaronsonGottesman}. In the following we will introduce an 
isomorphism (\ref{isomorphism}) between a factor of the Clifford group and the symplectic group $Sp_{2n}(2)$, 
which will be used to further analyze the structure and computational power of the Ising model TQC 
(cf. appendices \ref{app:B} and \ref{app:faith}). 

The Pauli group $\P_n$ for $n$ qubits is generated by the Pauli operators $\sigma_i$ acting on each of the  qubits. 
The elements of $\P_n$ can thus be written as tensor products 
of the Pauli matrices multiplied by overall phases, which are powers of $i$, 
\beq \label{Pauli}
\P_n=\left\{ i^{m}\sigma_{\alpha(1)} \otimes \cdots \otimes \sigma_{\alpha(n)}\, \left| \quad \alpha(j), \ m\in \{0,1,2,3\} \right.  \right\},
\eeq
where we have defined $\sigma_0=\I_2$. 
The Clifford group $\Cl_n$ for $n$ qubits is defined as the normalizer of the Pauli group, i.e.,
the group of those $n$-qubit unitary transformations $U$ which map the Pauli group $\P_n$ to itself under conjugation 
\beq
\label{DefClifford}
\Cl_n=\left\{ U \in SU(2^n)\ | \ U^*\cdot \P_n \cdot U \subset \P_n \ \right\}.
\eeq
Clearly, $\Cl_n$ is an infinite group since if $U$ satisfies Eq.~(\ref{DefClifford})
so does $\e^{i\phi}U$ for any $\phi\in \R$. 
However, the overall phases $\e^{i\phi}$ are in general irrelevant for quantum computation 
and therefore we could remove the center $Z$ of the Clifford group,
 which is generated by these phase factors. 
The resulting factor group 
\[
\P\Cl_n = \Cl_n / Z, 
\]
which we will call the \textit{projective Clifford group}, is already a finite group 
whose order will be derived later in this Section.
Notice that the minimal center of the Clifford group, which is required by the group 
multiplication law, is isomorphic to $\Z_8$ and is generated by $\e^{i\frac{\pi}{4}} \I$ as shown in 
Ref. \cite{MinimalPhase}.
The extension of the projective Clifford group by the center $\Z_8$ defines the minimal 
unprojected Clifford group which is also finite.

For $U\in \Cl_n$ we will denote the corresponding equivalence class in $\P\Cl_n$ by $[U]$. A frequently used 
discrete set of Clifford unitaries, whose images $[.]$ generate $\P\Cl_n$ if we allow them to act on arbitrary qubits is
\begin{equation}\label{CliffordGenerators}
P= \left[ \begin {array}{cc} 1&0\\
	\noalign{\medskip}0&i\end {array}
 \right]
,\quad
H= \frac{1}{\sqrt{2}}\left[ \begin {array}{cc} 1&1\\
	\noalign{\medskip}1&-1\end {array}   \right] \quad \mathrm{and}
\end{equation}
\[
 \mathrm{CNOT}=|0\ra\la 0|\otimes \I_2 + |1\ra\la 1|\otimes \sigma_1\,.
\]
It is obvious that the Pauli group is a subgroup of the Clifford group because 
\beq
\label{AntiCommutation}
\s_i^* \s_j \s_i  = \left\{\begin{array}{rr} \s_j &, i=j \\ -\s_j, & i\neq j \end{array}\right. 
\quad \Rightarrow \quad \P_n \subset \Cl_n .
\eeq

The equivalence classes of the elements of $\P_n$, which differ only by a phase, form a factor-group of $\P_n$, which we call the \textit{projective Pauli group}, 
\[
[\P_n]\equiv \P_n/\Z_4
\]
and it should play an important role in what follows.
Because of Eq.~(\ref{AntiCommutation}) the projective Pauli  group is also a normal subgroup of the projective Clifford group.

The main results obtained in this paper are based on the isomorphism between
the projective Clifford group,  factorized by its projective Pauli subgroup $[\P_n]$, and the symplectic group
$Sp_{2n}(2)$  over the finite field $\F_2$, i.e., 
\beq \label{isomorphism}
	\P\Cl_{n}/[\P_n] \simeq Sp_{2n}(2).
\eeq
Although this is a well known result we would like to explain it in more detail as it is of central importance for estimating the computational power of our topological quantum computer.
The group $Sp_{2n}(2)$ is defined as the set of invertible $2n\times 2n$ matrices with entries from the finite 
field $\F_2$ satisfying the relation
\beq
\label{SymplecticRelation}
S \in Sp_{2n}(2)\quad \Leftrightarrow \quad S^T \cdot M\cdot S = M
\eeq
where $M$ is the $2n\times 2n$  block-diagonal matrix
\[
M=\bigoplus^{n}\left[ \begin {array}{cc} \,\,\, 0&1 \\
	-1&0\end {array}
 \right] = \I_{n}\otimes 
\left[ \begin {array}{cc} \,\,\, 0&1 \\
	-1&0\end {array}  \right] .
\]
In order to understand the isomorphism (\ref{isomorphism}) we first note that the Pauli operators $\sigma_i$ ``commute'' up to phase factors. 
This means that, if we ignore these phase factors and keep track only of the multiplication rules, we obtain an additive group with  a ``modulo-2'' arithmetic, 
which is isomorphic to $\Z_2^2$. Indeed,  if we identify the Pauli matrices with the vectors
\beqa
\I \rightarrow (0,0)\quad &,& \quad \,\,\, \sigma_1 \rightarrow (1,0)\nonumber\\
\sigma_2 \rightarrow (0,1)\quad &,& \quad i\,\sigma_3 \rightarrow (1,1)\,\nonumber 
\eeqa
we could see that the Pauli multiplication rules are translated into vector summation over the finite field 
$\F_2$. 
In other words, the one-qubit projective Pauli group is isomorphic to the Abelian group
$\Z_2^2$
\[
\left[ \P_1\right] \simeq \left( \Z_2 \right)^2.
\]
This idea extends naturally to the $n$-qubit Pauli group $\P_n$, where we identify the $n$-qubit Pauli gates 
 with vectors in $\left(\F_2\right)^{2n}$, i.e., the vector space of the  $2n$-component vectors over the field $\F_2$, as illustrated by the example
\[
\sigma_1 \otimes \dots \otimes i\,\sigma_3 \rightarrow (1,0,\dots ,1,1)\, .
\]
Therefore we conclude that the projective $n$-qubit Pauli group is isomorphic to the Abelian group $\Z_2^{2n}$
\beq \label{P-order}
\left[ \P_n\right] \simeq \left(\Z_2^{2}\right)^{n}\equiv \Z_2^{2n}.
\eeq
Taking into account that the center of the $n$-qubit  Pauli group is $\Z_4$, generated by the powers of $i$, we could compute from Eq.~(\ref{P-order})  the order of the complete Pauli group  $\P_n$ to be
\beq\label{order-P_n}
|\P_n |=2^{2n+2}\,.
\eeq

Recall that the Clifford-group elements are structure preserving, invertible mappings (automorphisms) of the Pauli group, i.e. they map $\P_n$ to itself and respect products of Pauli operators. In more detail, if $U\in \Cl_n$ and we denote 
by $\underline{p}$ the multi-index of the Pauli operator $\s_{\underline{p}}\in \P_n$ then
\beq\label{U}
U^* \s_{\underline{p}} U = i^{f_U\left(\underline{p}\right)} \s_{S_U\left(\underline{p}\right)},
\eeq
where $f_U(\underline{p})$ is the power exponent of the phase, and we have denoted the mapping of the Pauli 
indices, corresponding to the action of the Clifford operator $U$ by
\beq\label{S_U}
S_U: \F_2 ^{2n}\to  \F_2 ^{2n}, \quad \underline{p} \to S_U\left(\underline{p}\right).
\eeq
While the index mapping (\ref{S_U}) essentially expresses the action of the Clifford gate $U$ inside of the projective Pauli group
$[\P_n]$, the pair $(f_U,S_U)$ in Eq.~(\ref{U}) describes unambiguously this action on the complete Pauli 
group $\P_n$.
The mapping (\ref{S_U}) is furthermore an automorphism of $\F_2^{2n}$  which could be characterized by the following properties:
\begin{enumerate}
\item \textbf{ $S_U$ is linear:}\\
It follows from the definition, Eq.~(\ref{U}), that
\[
\left[ \s_{S_U\left(\underline{p}\right)} \right]=\left[ U^* \s_{\underline{p}}U\right],
\]
where the brackets mean taking the projective equivalence class.
Therefore we can write, using the sign $\oplus$ for ``modulo-2'' addition of vectors,
\beqa
\left[ \s_{S_U\left(\underline{p}\oplus \underline{q}\right)} \right] &=&
\left[ U^* \s_{\underline{p}\oplus\underline{q}}U\right]=
\left[ U^* \s_{\underline{p}} \s_{\underline{q}}U\right] \nn
&=&
\left[ U^* \s_{\underline{p}} U\right] \left[U^*\s_{\underline{q}}U\right]=
\left[ \s_{S_U\left(\underline{p}\right)} \right] \left[ \s_{S_U\left(\underline{q}\right)} \right]\nn
&=&
\left[ \s_{S_U\left(\underline{p}\right)\oplus S_U\left(\underline{q}\right)} \right] \nonumber
\eeqa
from which we could derive the relation
\[
S_U\left(\underline{p}\oplus \underline{q}\right) = 
S_U\left(\underline{p}\right)\oplus S_U\left(\underline{q}\right),
\quad \forall  \underline{p},\ \underline{q}\in \F_2^{2n}.
\]
This means that $S_U\in GL_{2n}(2)$.
\item \textbf{$S_U$ is symplectic:}\\
Working with the projective Pauli group $[\P_n]=\F_2^{2n}$ is convenient, however, we loose the information about commutation or anti-commutation of the Pauli-operators, e.g. it is impossible to distinguish the two cases
\beqa
\sigma_2\cdot i\,\sigma_3 &=& \sigma_1 \nonumber \\
i\,\sigma_3\cdot \sigma_2 &=& -\sigma_1 \,.\nonumber
\eeqa
Nevertheless, it is possible to keep track of the corresponding commutation relations.
 If we define the inner product of two vectors $p=(p_1,p_2),\,q=(q_1,q_2) \in \F_2^2$ to be 
$p\ast q = p_2\cdot q_1 $ it is easy to check that
\[
 \sigma_p \cdot \sigma_q = (-1)^{p\ast q} \sigma_{p\oplus q}\,,
\]
where $\sigma_a$ denotes the Pauli-operator associated with $a\in \F_2^2$. This generalizes in a straightforward way to the $n$-qubit case where $\underline{p}=(p_1,\ldots,p_{2n})$,  $\underline{q}=(q_1,\ldots,q_{2n})$ and the inner product is given by
\[
\underline{p} \ast \underline{q} = \sum_{i=1}^n p_{2i}\cdot q_{2i-1}\,.
\]
Whether two operators $\sigma_{\p},\sigma_{\q} \in \P_n$ commute or not can be inferred from the symplectic form
\[
\omega\left(\p,\q\right)=\p\ast \q - \q\ast \p =\p^T\cdot M \cdot \q,
\]
which could only be equal to 0 or 1, where  $M$ is given in Eq.~(\ref{SymplecticRelation}) and  $\p^T$  is the transpose of $\p$.
It is easy to see that 
\beq\label{commute0}
\sigma_\p \cdot \sigma_\q = (-1)^{\omega\left(\p,\q\right)}\sigma_\q \cdot \sigma_\p\,, 
\eeq
from where we can conclude if the two operators $\s_\p$ and $\s_\q$  commute ($\omega(\p,\q)=0$)
or anti-commute ($\omega(\p,\q)=1$).

 Now we can prove that if the operator $U$  in Eq.~(\ref{U})  belongs to the Clifford group then
 $S_U$ must be symplectic, i.e., $\forall \,\p,\q\in \F_2^{2n}$
\beq\label{symp-S}
\omega\left(\p,\q\right)=
\omega\left(S_U\left(\p\right), S_U\left( \q\right)\right).
\eeq
Indeed, let us multiply Eq.~(\ref{commute0}) by $U^*$ from the left and by $U$ from the right.
We have
\[
U^* \sigma_\p U  U^* \sigma_\q U=  (-1)^{\omega\left(\p,\q\right)} U^* \sigma_\q U  U^* \sigma_\p U .
\]
Using Eq.~(\ref{U}) and canceling the phase factor $i^{f_U(\p)+f_U(\q)}$ on both sides we obtain
\beq\label{com-2}
\s_{S_U(\p)}\s_{S_U(\q)} =  (-1)^{\omega\left(\p,\q\right)}\s_{S_U(\q)}\s_{S_U(\p)}.
\eeq
On the other side, however, if $U\in \Cl_n$ as we assumed, then both $\s_{S_U(\p)}$ and $\s_{S_U(\q)}$ are Pauli operators as well, so they should also satisfy Eq.~(\ref{commute0}), i.e.,
\beq\label{com-3}
\s_{S_U(\p)}\s_{S_U(\q)} =  (-1)^{\omega\left(S_U(\p),S_U(\q)\right)}  \s_{S_U(\q)}\s_{S_U(\p)}.
\eeq
Comparing Eq.~(\ref{com-2}) with Eq.~(\ref{com-3}) we derive Eq.~(\ref{symp-S}) which implies that
$S_U$ is symplectic, i.e., 
\[
S_U^T\cdot M\cdot S_U=M\quad \Rightarrow \quad S_U \in Sp_{2n}(2).
\]
\item
\textbf{The kernel of $U\to S_U$ is the projective Pauli group $[\P_n]$:}\\
Consider the mapping between the Clifford and the symplectic groups
\[
S : \Cl_n \to Sp_{2n}(2),\, U \to S_U ,
\]
which is a homomorphism between the two groups $\Cl_n$ and $Sp_{2n}(2)$ both with matrix multiplication as group concatenation. The kernel of this homomorphism, i.e. the group of elements in $\Cl_n$ which are mapped to the identity in $Sp_{2n}(2)$ is generated by those  unitaries $U$ which satisfy
\beq
\label{Kernel}
U^* \s_{j_1}\otimes \dots \otimes \s_{j_n}U=\pm \s_{j_1}\otimes \dots \otimes \s_{j_n}
\eeq
for all choices of indices $j_1,\dots,j_n\in \{0,1,2,3\}$. Because of Eq.~(\ref{AntiCommutation}) all Pauli operators satisfy this equation, hence
\[
[\P_n]\subset \mathrm{Ker}\, S\,.
\]
On the other hand, since the Pauli operators form a basis for the linear operators on $\H^{\otimes n}$ we can 
express any unitary $U$ as a linear combination
\[
U=\sum_{\s_x \in \P_n}\alpha_x\s_x\,,\quad \alpha_x \in \C\,.
\]
Now suppose that there is a $U$ satisfying Eq.~(\ref{Kernel}) with more than one terms in the sum above. Then it is easy to see that there is a Pauli operator $\s_y$ such that the expansion of $U^*\s_yU$ consists of more than one Pauli operators which contradicts to the assumption that (\ref{Kernel}) is satisfied. 
Therefore the projected Pauli group is not only a subgroup of $\mathrm{Ker}\, S$, but also
\[
\mathrm{Ker}\,S \subset  [\P_n]\, \quad \Rightarrow \quad  \mathrm{Ker}\,S \equiv  [\P_n].
\]
\end{enumerate}
Given that the image of the mapping $S: U\to S_U$ is $Sp_{2n}(2)$ and that $\mathrm{Ker} \, S \equiv [\P_n]$
we conclude that
\[
  \P\Cl_n / \mathrm{Ker}\,S \simeq  \P\Cl_n /[\P_n]  \simeq Sp_{2n}(2)\,.
\]
This important fact allows us to use the results for the order of the  finite symplectic group
\beq \label{Sp-2}
	|Sp_{2n}(2)|  = 2^{n^2} \prod_{j=1}^n (4^j-1)\, ,
\eeq
which is proven in Appendix~\ref{app:A}, to derive the order of the projective Clifford group.
It follows from Eq.~(\ref{isomorphism}), and the fact that $[\P_n]$ has order $2^{2n}$, that
the order of the projective Clifford group is
\beq \label{PCl}
	|\P\Cl_{n}|=   2^{n^2+2n} \prod_{j=1}^n (4^j-1) \, .
\eeq
This result will be used in Sect.~\ref{sec:clifford-braid} to estimate the computational power of the Ising-anyon topological quantum computer.
\section{Braiding gates as Clifford gates: monodromy analysis}
\label{sec:braiding}
\begin{figure}[htb]
\centering
\includegraphics*[bb=10 300 580 540,width=8.5cm]{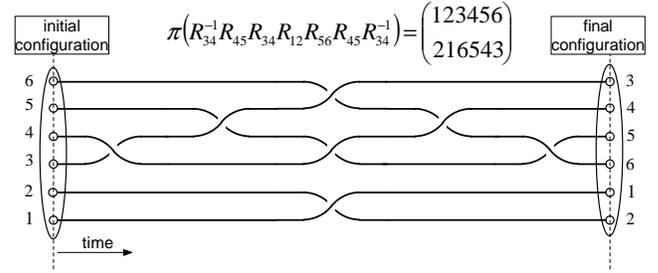}
\caption{Example of the projection mapping $\pi$ acting on the specific braid element
 CNOT$=R_{3}^{-1}R_{4}R_{3}R_{1}R_{5}R_{4}R_{3}^{-1} \in \B_6$, where $R_i\equiv R_{i,i+1}$ in the 
notation of Ref.~\onlinecite{TQC-NPB}, producing the permutation 	$(216543)\in \S_6$ . }
\label{fig:pi}
\end{figure}

\begin{figure}[htb]
\centering
\includegraphics*[bb=20 430 350 740,width=8.5cm]{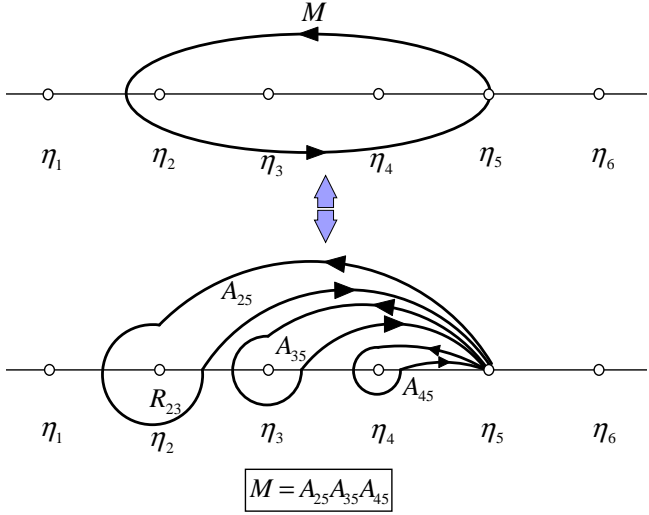}
\caption{Decomposition of a  monodromy transformation in which the particle with coordinate
	$\eta_5$ makes a complete loop around particles with coordinates $\eta_2$, $\eta_3$ and $\eta_4$,
	in terms of the monodromy generators $A_{ij}$. }
\label{fig:monodoromy}
\end{figure}

\begin{figure}[htb]
\centering
\includegraphics*[bb=20 430 350 740,width=8.5cm]{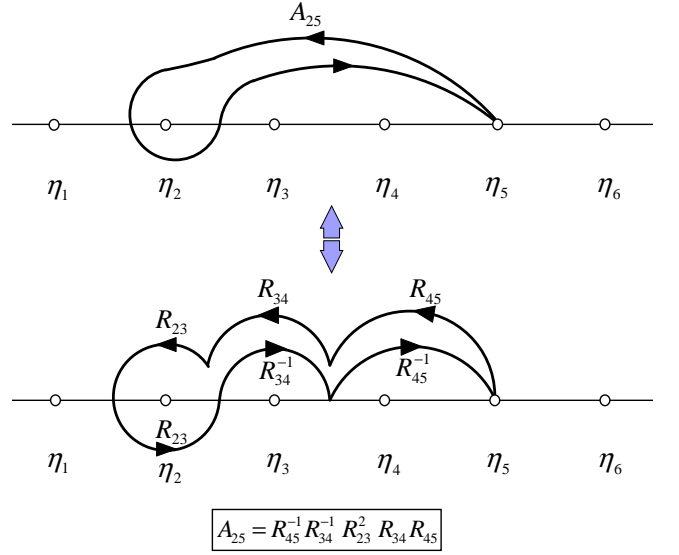}
\caption{The monodromy generator $A_{25}$, representing the elementary monodromy
	transformation in which the particle with coordinate $\eta_5$ is transported
	in a counter-clockwise direction along a complete loop around the particle with
	coordinate $\eta_2$, expressed in terms of the braid generators. }
\label{fig:A_25}
\end{figure}

A crucial observation for the results obtained in this paper, as well
as for topological quantum computation in general, is that the
Pauli group $\P_n$ for  $n$ qubits realized by $2n+2$ Ising anyons
coincides with the representation of the monodromy subgroup of the braid group
\[
	\P_n \equiv \mathrm{Image}\left(\M_{2n+2} \right)
\]
Because the monodromy group is in general a normal subgroup of the braid group
it follows that all braiding gates, which by definition leave the monodromy
(hence the Pauli group) invariant, are actually Clifford gates, i.e.,
 \[
	 \mathrm{Image}(\B_{2n+2}) \subset \Cl_n    .
 \]
In order to give a more comprehensive explanation of this important result
we shall recall the standard algebraic
definition of the monodromy subgroup in terms of the exact sequence of
group homomorphisms
 \[
	 1 \to \M_n \to \B_n \to \S_n \to 1
 \]
where $\M_n$ is the monodromy group (also called the pure braid group),
$\B_n$ is the full braid group and $\S_n$ is the symmetric group.
In more detail, recall that a braid can be described geometrically
as a bunch of strands connecting the initial configuration of some indistinguishable
particles (arranged for simplicity along a line) with the final configuration.
If we forget about the strands and look at the final positions of the particles,
arranged along the same line, then the final configuration will be simply a
permutation of the original particles. As an illustration
we show in Fig.~\ref{fig:pi} the action of the projection map on the two-qubit
CNOT gate of Refs.~\onlinecite{TQC-PRL,TQC-NPB}, which is an element of the Ising-model
representation of $\B_6$.
Therefore we can define the natural
``onto''- mapping
\[
\pi : \B_n \to \S_n,
\]
which could be easily seen to be a group-homomorphism. Next, as is well known, the
kernel of any group homomorphism is a normal subgroup and the factor-group
is isomorphic to the symmetric group
\[
\B_n/\mathrm{Ker} \, \pi \simeq \S_n.
\]
Therefore we may define the monodromy group  as   the kernel of the
projection $\pi$, i.e.,
\[
\M_n \equiv \mathrm{Ker} \, \pi .
\]
Geometrically this definition means that a monodromy transformation is any braid transformation
which does not permute the final positions of the particles compared to the initial ones.
Pictorially this could be described by such braid transformations in which one or more
particles are transported along closed loops around one or more other particles
(so that each particle arrives at the end at its original position)
as shown in Fig.~\ref{fig:monodoromy}.
It is not difficult to see that these pure braid transformations can be generated by
the elementary monodromy transformations in which one particle, say with index $j$, is
transported along a complete loop around another particle, say with index $i$, where $i<j$.
Thus the elementary generators of the monodromy group could be given the following 
presentation \cite{Birman}
\beqa \label{A_ij}
A_{ij}&=&  R_{j-1}^{-1} R_{j-2}^{-1} \cdots  R_{i+2}^{-1} R_{i+1}^{-1}\ R_{i}^2 \ R_{i+1} R_{i+2} \cdots R_{j-2} R_{j-1} \nn
&=& U_{ij}^{-1} R_{i}^2 U_{ij} , \quad \mathrm{where} \quad U_{ij}= \prod_{k=i+1}^{j-1} R_k,
\eeqa
and $1\leq i<j \leq n$.
In other words the monodromy group could be considered as the smallest invariant subgroup of
$\B_n$ which is generated by the squares of the elementary braid
generators \cite{todorov-stanev} $R_i^2$, $i=1,\ldots, n-1$.
As an illustration of Eq.~(\ref{A_ij})
we display in  Fig.~\ref{fig:A_25} the monodromy transformation in which the
particle with coordinate $\eta_5$ is transported along a complete counter-clockwise
loop around the particle with coordinate $\eta_2$.

Alternatively, the monodromy group could be  abstractly defined
by the generators $A_{ij}$, $1\leq i<j \leq n$ satisfying certain relations,
however we will skip that as we shall only need here the expressions (\ref{A_ij})
in terms of the braid generators.

It can be seen that in the braid-group representation of the Ising model
the Pauli group is represented in terms of monodromy operators. Indeed,
the $\sigma_3$ gate acting on the $i$-th qubit is simply
\beq \label{square-1}
\left(R_{2i-1}^{(n+1,+)} \right)^2 =
\underbrace{\I_2\otimes  \cdots \otimes \I_2}_{i-1} \otimes \sigma_3 \otimes
\underbrace{\I_2\otimes \cdots \otimes \I_2}_{n-i},
\eeq
while the squares of the non-diagonal braid generators give
\beq \label{square-2}
\left(R_{2i}^{(n+1,+)} \right)^2 =
\underbrace{\I_2\otimes \cdots \otimes \I_2}_{i-1} \otimes \sigma_2\otimes \sigma_2 \otimes
\underbrace{\I_2\otimes  \cdots \otimes \I_2}_{n-i-1},
\eeq
where $1\leq i\leq n-1$.
The last two squares are more special:
\beq \label{square-3}
\left(R_{2n}^{(n+1,\pm)} \right)^2 =
\mp\underbrace{\sigma_3\otimes \cdots \otimes \sigma_3}_{n-1} \otimes \sigma_1
\eeq
\beq \label{square-4}
\left(R_{2n+1}^{(n+1,\pm)} \right)^2 = \pm
\underbrace{\sigma_3\otimes \cdots \otimes \sigma_3}_{n}          .
\eeq
It is obvious from the above equations
that all $n$-qubit Pauli gates  can be expressed (up to phases which are
powers of $i$) in terms of
products of squares of the elementary braid generators.
For example, the $\sigma_2$ gate acting on the last qubit is
\[
\sigma_2^{(n)} = i \left(R_{2n}^{(n+1,+)}\right)^2 \left(R_{2n+1}^{(n+1,+)}\right)^2\, ,
\]
and the $\sigma_2$ gate acting on qubit $(n-j)$ can be expressed as
\[
\sigma_2^{(n-j)} = i \left(R_{2n-2j}^{(n+1,+)}\right)^2 \s_2^{(n-j+1)} , \quad
1\leq j\leq n-1.
\]
The above equations also suggest that we might need some additional phase factors in order
to construct all elements of $\P_n$.
To clarify this point we will construct the element $i\I$ explicitly in the positive-parity representation 
(a similar construction can also be given for the negative-parity representation because both representations 
are computationally equivalent, as shown in Ref.~\onlinecite{equiv}). More precisely, we will prove below that
\beq
\label{PhaseI}
R_{2n}^{(n+1,+)}\left(R_{2n+1}^{(n+1,+)}\right)^2 R_{2n}^{(n+1,+)}\left(R_{2n+1}^{(n+1,+)}\right)^2=i\, \I\,.
\eeq
The left hand side corresponds to a composition of two braids, where particle $2n$ moves along a complete 
loop around particles $2n+1$ and $2n+2$ and likewise particle $2n+1$ moves around $2n+2$. 
Therefore Eq.~(\ref{PhaseI}) represents an element of the monodromy subgroup and we conclude that
\[
\{\pm \I,\pm i \I\}\subset \textrm{Image}\left(\M_{2n+2}\right).
\]
To verify (\ref{PhaseI}) we first observe that the last two unprojected braid matrices
$R_{2n}^{(n+1)}$ and $R_{2n+1}^{(n+1)}$ satisfy, according to Eq.~(\ref{R}) and the anticommutation 
relations of the $\gamma$ matrices, the following relation
\beq \label{i}
R_{2n}^{(n+1)}\, \left(R_{2n+1}^{(n+1)}\right)^2 \, R_{2n}=
i \, \left(R_{2n+1}^{(n+1)}\right)^2\,.
\eeq
Indeed, using the properties of the $\gamma$ matrices, it is easy to see that the
squares of the unprojected braid matrices (\ref{R}) are simply
\beq
\label{squares}
\left(R_{j}^{(n+1)}\right)^2 = -i\,\gamma_{j}^{(n+1)} \gamma_{j+1}^{(n+1)},  \quad
1\leq j \leq n-1.
\eeq
Applying Eq.~(\ref{squares}) for $j=2n+1$, and using the anticommutation relations (\ref{cr})
for the $\gamma$ matrices, 
 we have for the LHS of  Eq.~(\ref{i})
\beqa
\mathrm{LHS}&=&\frac{i}{2}
\left(\I -\gamma_{2n}\gamma_{2n+1} \right)\left(-i\gamma_{2n+1}\gamma_{2n+2} \right)
\left(\I -\gamma_{2n}\gamma_{2n+1} \right)  \nn
&=&i(-i)\gamma_{2n+1}\gamma_{2n+2} = i \, \left(R_{2n+1}\right)^2. \nonumber
\eeqa
The fourth power of the unprojected braid matrices is always $\I$ because
applying twice Eq.~(\ref{squares}) we have
\[
\left(R_{j}^{(n+1)}\right)^4 = 
-\gamma_{j}^{(n+1)}\gamma_{j+1}^{(n+1)}\gamma_{j}^{(n+1)}\gamma_{j+1}^{(n+1)}=\I ,
\]
which also implies that the fourth power of the projected braid generators (\ref{R_pm})
is $\I$. Finally, the validity of Eq.~(\ref{PhaseI}) follows from the fact that the
projectors (\ref{P_pm}) commute with the group multiplication law (see Eq.~(\ref{commute}) below).

Now, because the squares $\left( R_{i}^{(n+1,\pm)}\right)^2$ of the elementary braid
generators, as well as the element $i\I$, belong to the monodromy group representation, we conclude that the 
Pauli group $\P_n$ is naturally represented in the Ising model TQC as a subgroup of the
monodromy group for $2n+2$ Ising anyons, i.e. ,
\beq \label{P_n-in-M}
	\P_n \subset \textrm{Image}\left(\M_{2n+2}\right).
\eeq
On the other hand, as we will show below, the generators (\ref{A_ij}) of the
monodromy group could be entirely expressed in terms of the Pauli operations
for $n$ Ising qubits so that
\beq \label{mono-pauli}
\textrm{Image}\left( \M_{2n+2} \right) \subset	\P_n \ \  \Rightarrow  \ \    \P_n \equiv 
\textrm{Image}\left(\M_{2n+2} \right).
\eeq
In order to prove the group isomorphism  (\ref{mono-pauli}) we will show that the monodromy
generators of the Ising-model representation of $\M_{2n+2}$ could be expressed (up to powers of $i$) in 
terms of the squares of the elementary braiding matrices $\left(R_{i}^{(n+1,+)} \right)^2$, which, according to 
Eqs.~(\ref{square-1}), (\ref{square-2}), (\ref{square-3}) and  (\ref{square-4}), are completely expressible in 
terms of Pauli operators.

Before we prove this let us emphasize once again one important detail of the Ising representation of the braid group. 
The projectors (\ref{P_pm}) commute with the group multiplication law, i.e.,
\beq \label{commute}
R_{i}^{(n+1,\pm)}R_{j}^{(n+1,\pm)} =
R_{i}^{(n+1)}R_{j}^{(n+1)} P_{\pm}^{(n+1)}, 
\eeq
so that, if we want, we could work all the time with the unprojected matrices and project them at the end. 

Using Eq.~(\ref{squares}) as a base we could prove by induction that the unprojected monodromy generators  are
\beq \label{A_kl-un}
A_{kl}=i (-1)^{l-k} \gamma_{k}^{(n+1)} \gamma_{l}^{(n+1)}, \quad 1\leq k < l \leq 2n+2.
\eeq
Indeed, assume that Eq.~(\ref{A_kl-un}) is fulfilled for given $k$ and $l$ where $l>k$. Then
\beqa
A_{k,l+1}&=&R_l^{-1} A_{kl} R_l \nn
&=& \frac{1}{2}
\left(\I +\gamma_{l}\gamma_{l+1} \right)\left(i (-1)^{l-k}\gamma_{k}\gamma_{l} \right)
\left(\I -\gamma_{l}\gamma_{l+1} \right),\nonumber
\eeqa
where we have used that the inverse of the unprojected braid generator $R_l$ is $R_{l}^{-1} = \e^{-i\pi/4}
\left(\I +\gamma_{l}\gamma_{l+1} \right)/\sqrt{2} $. Then using the anticommutation relations
(\ref{cr}), to simplify expressions such as
\[
\gamma_l\gamma_{l+1}\gamma_k\gamma_l\gamma_l\gamma_{l+1}=
\gamma_l\gamma_{l+1}\gamma_k\gamma_{l+1}=-\gamma_l\gamma_k,
\]
we obtain $A_{k,l+1}=-i (-1)^{l-k}\gamma_k\gamma_{l+1}$ which completes the proof of Eq. (\ref{A_kl-un}).
Next, plugging between $\gamma_k$ and $\gamma_l$ the identities $\gamma_{k+1}\gamma_{k+1}$,
$\gamma_{k+2}\gamma_{k+2}$ until $\gamma_{l-1}\gamma_{l-1}$
and re-expressing $\gamma_k\gamma_{k+1}=i \left(R_k\right)^2$ we can write the unprojected monodromy generators as
\[
A_{kl}= - (-i)^{l-k+1} \left(R_k\right)^2\left(R_{k+1}\right)^2 \cdots \left(R_{l-2}\right)^2\left(R_{l-1}\right)^2.
\]
Finally, using the fact that projectors commute with the group multiplication,  as in Eq.~(\ref{commute}), we 
obtain e.g., the monodromy generators in the positive-parity representations (and similar for negative parity) as
\beq\label{A_kl}
A_{kl}^{+}= - (-i)^{l-k+1} \left(R_k^{+}\right)^2\left(R_{k+1}^{+}\right)^2 \cdots \left(R_{l-2}^{+}\right)^2\left(R_{l-1}^{+}\right)^2,
\eeq
where $1\leq k < l \leq 2n+2$ and
$R_k^{+}= R_{k}^{(n+1)}P_{+}^{(n+1)}$.
Going back to Eqs.~(\ref{square-1}), (\ref{square-2}), (\ref{square-3}) and (\ref{square-4})
we conclude that because $\left(R_k^{+}\right)^2 \in \P_n$ for all $k$ and the element $i\I \in \P_n$  as well (see Eq.~(\ref{PhaseI})), the generators of the monodromy representation
certainly belong to the Pauli group  $\P_n$ so that
\beq \label{A_kl-plus}
A_{kl}^{+} \in \P_n \quad \Rightarrow \quad
\textrm{Image}\left(\M_{2n+2} \right)\subset \P_n.
\eeq
Combining Eq.~(\ref{A_kl-plus}) with Eq.~(\ref{P_n-in-M}) we finally prove that
\textit{the Ising-model monodromy group representation of $\M_{2n+2}$ is completely equivalent
 to the $n$-qubit Pauli group $\P_n$ for the Ising-model topological quantum computer.} \\

\noindent
\textbf{Remark 1:} \textit{ It follows from Eq.~(\ref{mono-pauli}) that the center of the monodromy group representation
coincides with the center of the $n$-qubit Pauli group which is generated by $i\I$, i.e.,
$\textrm{Center}\left(\textrm{Image} (\M_{2n+2}) \right) = \textrm{Center}\left(\P_n\right) = \Z_4$.
Combining this with the algebraic fact \cite{Birman} that the permutation group $\S_{2n+2}$ is centerless for $n\geq 1$,
so that the center of the braid group coincides with the center of its monodromy subgroup,
and with the fact that  $\S_{2n+2}$ is represented faithfully in the Ising-model representation of $\B_{2n+2}$ for 
$n\geq 2$, as proven in Appendix~\ref{app:faith}, we conclude that
}
\beqa\label{center}
\textrm{Center}\left(\textrm{Image} \left(\M_{2n+2}\right) \right) &\equiv &
\textrm{Center}\left(\textrm{Image} \left(\B_{2n+2}\right) \right) \nn
&=& \left\{\pm \I,\pm i \I \right\}, \quad n \geq 2.
\eeqa

\noindent
\textbf{Remark 2:} \textit{Because the monodromy group $\M_{2n+2}$ in the Ising-model representation of the 
braid group $\B_{2n+2}$ exactly coincides with the $n$-qubit Pauli group $\P_n$ this automatically means that the 
image of the braid group is a subgroup of the $n$-qubit Clifford group $\Cl_n$
\[
\mathrm{Image} \left( \B_{2n+2}\right) \subset \Cl_n ,
\]
i.e., all quantum gates that could be implemented by braiding in the Ising-model TQC are in fact Clifford gates 
that stabilize the Pauli group. Unfortunately, the converse is not true,
i.e.,  it appears that not all Clifford gates could be implemented by braiding Ising anyons.
}

Furthermore, we could use the group isomorphism (\ref{mono-pauli}) between the monodromy group 
$\textrm{Image}\left(\M_{2n+2}\right)$ and the Pauli group $\P_n$
to derive the order of the image of the braid group $\B_{2n+2}$.
Recall that in general the factor-group $\B_{2n+2}/\M_{2n+2}$ is isomorphic to the permutation group $\S_{2n+2}$. 
In our case it appears that for $n\geq 2$ this permutation group is represented faithfully (see App.~\ref{app:faith} 
for a proof) so that the order of the Ising-model braid-group representation is
\beqa \label{order}
\left| \mathrm{Image}\left( \B_{2n+2}\right)\right|
&=&  \left| \mathrm{Image}\left( \M_{2n+2}\right)\right|  \left| \mathrm{Image}\left( \S_{2n+2}\right)\right| \nn
&=& 2^{2n+2}  (2n+2)!, \quad n\geq 2,
\eeqa
where we have used that because of Eq.~(\ref{mono-pauli}) the order of the monodromy group
is equal to the order of the full $n$-qubit Pauli group $\P_n$ (including the center, cf. Eq.~(\ref{center})), 
which is $2^{2n+2}$ and that the order of the faithful representation of $\S_{2n+2}$ is $(2n+2)!$.
For $n=1$ the order of the image of the monodromy group is again given by the order of the Pauli group $2^{2n+2}=16$, however, the representation of the permutation group $\S_4$ is not faithful. Instead, the third generator is equivalent to the first one, i.e., $R_{3}^{(+)} = R_{1}^{(+)}$ or
$R_{3}^{(-)} = i \left(R_{1}^{(-)}\right)^{-1}$ and therefore the factor-group gives a faithful representation of the permutation group $\S_3$. Thus the image of the braid group for 4 Ising anyons is
\beqa
\left| \mathrm{Image}\left( \B_{4}\right)\right|
&=&  \left| \mathrm{Image}\left( \M_{4}\right)\right|  \left| \mathrm{Image}\left( \S_{3}\right)\right| \nn
&=&16\cdot3!=96, \nonumber
\eeqa
which exactly coincides with the result of Ref.~\onlinecite{TQC-NPB} where the order has been directly computed using the Dimino's algorithm. The result, $2^{6}6!=46080$, for the order of the image of the braid group $\B_{6}$ computed from Eq.~(\ref{order}) also exactly matches that
obtained by the Dimino's algorithm in Ref.~\onlinecite{TQC-NPB}.

Finding the order of the image of the braid group $\B_{2n+2}$
is a central result in this paper which is similar to the results of Ref.~\onlinecite{read-JMP}, where the order of the 
image of the braid group depends on the parity of $n$, yet it is different. The reason for the independence of the 
order of the representation on the parity of $n$ in our case is the presence of the additional phase 
factor $\e^{i\frac{\pi}{4}}$ in Eq.~(\ref{R}), which certainly changes the center of the representation of the 
braid group.
\section{The projective Clifford group $\P\Cl_n$ and the Ising representation
of the braid group $\B_{2n+2}$}
\label{sec:clifford-braid}
In this Section we shall prove that not all Clifford-group gates
could actually be implemented by braiding of Ising anyons. To this end we shall demonstrate
that the order of the projective Clifford group, i.e., the Clifford group factorized by
its center,  is much bigger and grows much faster with $n$ than the order of the image of the braid group. 
Because global phases are irrelevant for quantum computation it makes sense
to compare  the orders of the two finite groups factorized by their centers.

Let us first compare the single-qubit braid group $\B_4$ and the single-qubit Clifford group $\P\Cl_1$.
The order of the Ising-model representation of $\B_4$ is 96, however if we factorize it by its $\Z_4$ center
we have $\left|\mathrm{Image}\left(\B_4\right)/ \Z_4 \right|=24$. This precisely coincides with the order of the projective Clifford group for one qubit obtained from Eq.~(\ref{PCl}).

\begin{figure}[htb]
\centering
\includegraphics*[bb=55 30 580 440,width=8cm]{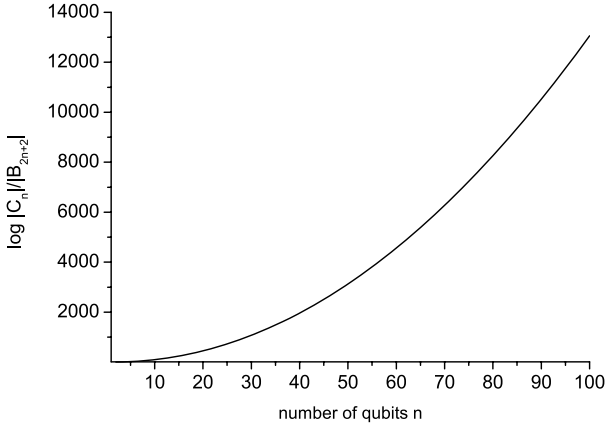}
\caption{Comparing the order of the $n$-qubit Clifford group $\Cl_n$  with that of the image of the $n$-qubit  braid group $\B_{2n+2}$. The logarithm of the ratio of the two orders is plotted on the vertical axis versus the number of qubits $n$. }
\label{fig:log}
\end{figure}

For two qubits the orders of the representation of the braid group $\B_6$  and its projective version are
\[
 |\mathrm{Image}(\B_{6})| = 46080   \quad \Rightarrow  \quad
 |\mathrm{Image}(\B_{6})/\Z_4| =11520
\]
and again the second number precisely coincides with the order of the two-qubit projective Clifford group 
obtained from Eq.~(\ref{PCl})
\[
	|\P\Cl_{2}|= 2^{2^2+4} (4-1)(4^2-1) =  11520.
\]
While the coincidence of the orders of the one-qubit braid group representation and the one-qubit Clifford group seems 
natural, the one for the two-qubit groups is more subtle. It  could be understood by the occasional group isomorphism 
\cite{planat-clifford,wilson}
\[
	\P\Cl_{2} /[\P_2]\simeq \S_6   \equiv \B_6/ \M_6  .
\]
and this remarkable mathematical fact could  eventually explain why the entire Clifford group for
two qubits can indeed be realized by braiding, while for any $n\geq 3$ the Clifford group
is much bigger, and grows much faster with $n$, than the image of the braid group.
Indeed, the orders of the image of the braid group (factored by its center) and of the projective Clifford group
for $n=1, \ldots 5$ number of qubits are compared in Table~\ref{tab:compare}. Already for 3 qubits the order of the projective Clifford group is about 40 times bigger than the order of the image of the braid group factored by its center.
\begin{table}[htb]
\caption{Comparing the orders of the image of the braid group $\B_{2n+2}$,
factorized by its center $\Z_4$, and of the projective Clifford group $\P\Cl_n$. \label{tab:compare}}
 \begin{tabular}{l||r|r|r|r|r}
 	\hline
	$\#$ of qubits	$n$ & 1 & 2 & 3 & 4 & 5  \\
\hline \hline
$\left|\mathrm{Im}(\B_{2n+2})/\mathrm{\Z_4}\right|$& 24 & 11520 & 2580480 & $\propto 0.9\times 10^{9}$
& $\propto 0.5\times 10^{12}$ \\
\hline
$|\P\Cl_n|$ & 24 &  11520 & 92897280 & $\propto 1.2\times 10^{13}$ &
$\propto 2.5\times 10^{19}$\\
\hline
	\end{tabular}
\end{table}
For $n=4$ and $5$ the discrepancy is even more dramatic: the order of the projective Clifford group is increasing much faster with $n$ than the order of the image of the braid group. In Fig.~\ref{fig:log} we have plotted the logarithm of the ratio between the order of the projective Clifford group   and the order of the braid group for number of qubits up to 100.
To understand more intuitively  Fig.~\ref{fig:log} notice that the 
leading term in the order of the projective Clifford group is 
\[
|\P\Cl_n| \simeq 2^{n^2+2n} \prod_{j=1}^n 2^{2j} \propto 2^{2n^2+3n},
\]
while (using the Stirling formula) the order of the image of the (projective) braid group is
\[
|\mathrm{Image}(\B_{2n+2})|=2^{2n} (2n+2)! \propto 2^{2n +(2n+2)\log_2(2n+2)}
\]
and therefore  the 
leading term in the ratio is 
\[
|\P\Cl_n| /  |\mathrm{Image}\B_{2n+2}|\propto 2^{2n^2+n - (2n+2)\log_2(2n+2)} \propto 2^{2n^2} ,
\]
 so that
the logarithm of this ratio still grows quadratically with $n$.

It is not difficult to see which gates are missing, i.e., which are the Clifford gates that cannot be implemented by braiding of Ising anyons. These are the embeddings of the two-qubit SWAP gate into the $n$-qubit system.
Indeed, if we have them then we could construct by braiding all embeddings of the CZ gates 
because by physically exchanging pairs of Ising anyons, representing the qubits, we readily  obtain CZ.SWAP
as illustrated in Eq.~(\ref{CZ.SWAP}) below for the two-qubit case. 
In general, by exchanging the $j$-th and $(j+1)$-th pair of anyons, with coordinates
$(\eta_{2j-1},\eta_{2j})$ and $(\eta_{2j+1},\eta_{2j+2})$ respectively, 
 we obtain (cf. (\ref{CZ.SWAP}))
\beqa
R^{(n+1,+)}_{2j}R^{(n+1,+)}_{2j+1}R^{(n+1,+)}_{2j-1}R^{(n+1,+)}_{2j}= \nn
=i\mathrm{CZ}^{(j,j+1)}.\mathrm{SWAP}^{(j,j+1)}, \nonumber
\eeqa
where $\mathrm{CZ}^{(j,j+1)}$ is  the Control-Z gate acting on the $j$-th qubit (the control qubit) and $(j+1)$-th qubit
(the target qubit) and $\mathrm{SWAP}^{(j,j+1)}$ is the two-qubit SWAP gate exchanging the  $j$-th and $(j+1)$-th 
qubits.
The Phase gate $P$ in (\ref{CliffordGenerators}) acting on qubit $j$ is given by $R_{2j-1}^{(n+1,+)}$. 
Moreover, if we have all the SWAPs then we can construct Hadamard gate acting on arbitrary qubit by swapping the Hadamard gate
\[
H^{(n)}= \left( R^{(n+1,+)}_{2n-1} \right)^2 R^{(n+1,+)}_{2n+1} R^{(n+1,+)}_{2n} \left( R^{(n+1,+)}_{2n+1} \right)^{-1}
\]
acting on the last qubit to the qubit where it is needed.

Thus we see that if we could construct all embeddings of the two-qubit SWAP gate then we could construct the entire Clifford group for $n$ qubits.
\section{Solution for two qubits: braiding implementation of the SWAP gate}
\label{sec:2qubits}
\begin{figure}[htb]
\centering
\includegraphics*[bb=15 350 580 490,width=8cm]{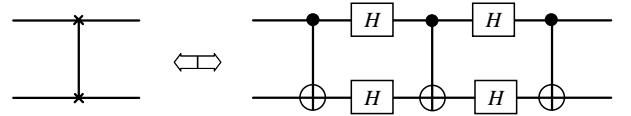}
\caption{SWAP gate in terms of 3 CNOT and 4 Hadamard gates }
\label{fig:SWAP}
\end{figure}

\begin{figure}[htb]
\centering
\includegraphics*[bb=30 590 300 680,width=7cm]{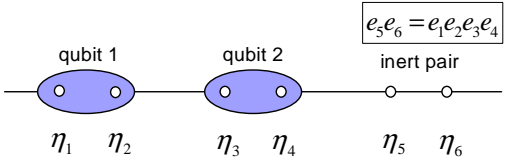}
\caption{Two-qubits configurations in terms of 6 Ising quasiholes corresponding  to the positive-parity 
representation of the braid group $\B_6$.}
\label{fig:2qubits}
\end{figure}

The problem with the implementation of all Clifford gates could be most easily
illustrated on the example of the two-qubit SWAP gate \footnote{we thank Sergey
Bravyi for pointing this fact to us}, which is a Clifford gate that can be expressed
in terms of CNOT and Hadamard gates \cite{nielsen-chuang} as shown in Fig.~\ref{fig:SWAP}.
Imagine a two-qubit system constructed from 6
Ising anyons, with coordinates $\eta_1,\ldots,\eta_6$ in the plane, in such a way
that the first pair (with coordinates $\eta_1$ and $\eta_2$) forms the first qubit,
the second pair (with coordinates $\eta_3$ and $\eta_4$) - the second qubit and
the last two qubits with coordinates $\eta_5$ and $\eta_6$ form the inert pair
as shown in Fig.~\ref{fig:2qubits}.
Without restriction of generality we could choose to work in the spinor representation
$S_+$ with positive total parity because, as shown in Ref.~\cite{equiv},
both inequivalent spinor representations $S_\pm$ of the braid group $\B_{2n+2}$ are computationally equivalent.
The two-qubit computational basis in the spinor representation $S_+$ can be written
in terms of Ising-model correlation functions as
\beqa \label{2qb-basis1}
|00\ra &\equiv& \la \s_+\s_+ \ \s_+\s_+ \ \s_+\s_+ \ra \nn
|01\ra &\equiv& \la \s_+\s_+ \ \s_+\s_- \ \s_+\s_- \ra \nn
|10\ra &\equiv& \la \s_+\s_- \ \s_+\s_+ \ \s_+\s_- \ra \nn
|11\ra &\equiv& \la \s_+\s_- \ \s_+\s_- \ \s_+\s_+ \ra  .
\eeqa
Now, let us try to swap the two qubits by simply exchanging the two pairs
comprising the qubits. This could be done by the following braid transformation
(see Ref.~\cite{equiv} for the explicit form of the braid matrices in this basis)
\beq \label{CZ.SWAP}
U^{(2)}_{\# 1 \leftrightarrow  \# 2} =
R^{(3,+)}_{2}R^{(3,+)}_{3}R^{(3,+)}_{1}R^{(3,+)}_{2} = i
\left[ \begin {array}{cccc} 
1&0&0&0\\
0&0&1&0\\
0&1&0& 0\\
0&0&0&\!\!\!-1\end {array}
 \right].
\eeq
The resulting unitary transformation has one extra minus on the last row
as compared to the standard SWAP gate \cite{nielsen-chuang} (we skip the global phase $i$).
This is, however, natural because this minus sign
appears in front of the state $|11\ra$ that is the state in which both Ising
anyon pairs at $(\eta_1, \eta_2)$ and at $(\eta_3, \eta_4)$ share Majorana fermions
and therefore  exchanging the two qubits is equivalent (especially after fusing
$\eta_1 \to \eta_2$  and  $\eta_3 \to \eta_4$) to the exchange of two fermions
which certainly produces a minus sign.
This raises the reasonable question whether the SWAP gates is at all realizable
by braiding.

\begin{figure}[htb]
\centering
\includegraphics*[bb=30 590 300 680,width=7cm]{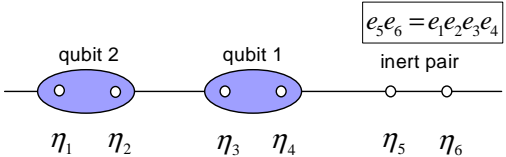}
\caption{Two-qubits configurations corresponding  to the positive parity, however, with
different encoding of the qubits. }
\label{fig:2qubits2}
\end{figure}

On the other hand, we could have chosen a different basis of Ising $\s$ fields
to represent two qubits, e.g.,
the first pair (with coordinates $\eta_1$ and $\eta_2$) forms the second qubit,
while the second pair (with coordinates $\eta_3$ and $\eta_4$) forms the first qubit and
the last two qubits with coordinates $\eta_5$ and $\eta_6$ form the inert pair again
as shown in Fig.~\ref{fig:2qubits2}.
Then the two-qubit computational basis can be written
in terms of Ising-model correlation functions as
\beqa \label{2qb-basis2}
|00\ra' &\equiv& \la \s_+\s_+ \ \s_+\s_+ \ \s_+\s_+ \ra  \nn
|01\ra' &\equiv& \la \s_+\s_- \ \s_+\s_+ \ \s_+\s_- \ra \nn
|10\ra' &\equiv& \la \s_+\s_+ \ \s_+\s_- \ \s_+\s_- \ra \nn
|11\ra' &\equiv& \la \s_+\s_- \ \s_+\s_- \ \s_+\s_+ \ra  .
\eeqa
Obviously the basis (\ref{2qb-basis2}) could be obtained from the basis
(\ref{2qb-basis1}) by the action of the transformation matrix
\beq\label{SWAP}
A^{(2)} =
\left[ \begin {array}{cccc} 
1&0&0&0\\
0&0&1&0\\
0&1&0& 0\\
0&0&0&1\end {array}
 \right] \equiv \mathrm{SWAP}.
\eeq
which simply means that the bases (\ref{2qb-basis1}) and
 (\ref{2qb-basis2}) describe equivalent representations
of the braid group $\B_6$.

The question whether the SWAP gate is implementable by braiding is equivalent to
that if the product
\[
U_{\# 1 \leftrightarrow  \# 2} \ A^{(2)}=i \ \mathrm{diag}(1,1,1,-1),
\]
which is (up to phase) equal to the Controlled-$Z$ gate, is expressible in terms
of the braid-group generators. Indeed, using the explicit form of the generators from
Ref.~\onlinecite{equiv}, we could easily check that
\[
R^{(3,+)}_{1}R^{(3,+)}_{3}\left(R^{(3,+)}_{5}\right)^{-1}=\mathrm{diag}(1,1,1,-1)=
\mathrm{CZ},
\]
which constructively proves that the two-qubit CZ and SWAP gates (hence, all Clifford
gates for two qubits) are efficiently realizable by braiding of Ising anyons.
\section{Conclusion}
In this paper we have demonstrated that the Pauli group in the Ising-anyon
topological quantum computer exactly coincides with the image of the monodromy group.
Therefore,  all quantum gates that can be implemented by braiding of Ising anyons are Clifford gates 
that stabilize the Pauli group. Analyzing the structure of the Clifford
group for $n$-qubits and comparing to the structure of the braid-group representation
of the Ising model, we conclude that not all Clifford gates could in general be realized by
braiding in this model. The gates that are missing are the embeddings of the two-qubit SWAP gate
into a $n$-qubit Ising systems.
Thus, not only the Ising-anyon topological quantum computer is not universal for topological
quantum computation but even the entire Clifford group is not implementable by braiding.
Nevertheless, the braid-group representation of the Ising TQC contains all of the entangling gates that could be 
used to analyze topological entanglement and topological protection of qubits and quantum operations.
\begin{acknowledgments}
We thank Sergey Bravyi, Holger Vogts, Lyudmil Hadjiivanov,  Volkher Scholz and Johannes Guetschow
for useful discussions. 
A.A. acknowledges support from "DFG Forschergruppe 635".
L.S.G. thanks the Galileo Galilei Institute for Theoretical Physics in Firenze, 
Italy, for hospitality as well as INFN for partial support. L.S.G. has been supported as a 
Research Fellow by the Alexander von Humboldt foundation and by the BG-NCSR under
Contract No. F-1406.
\end{acknowledgments}
\begin{appendix}

\section{Order of the symplectic group}
\label{app:A}
Following Ref.~\cite{wilson} we could find the order of the symplectic group
$Sp_{2n}(q)$ over the field $\F_q$ by counting the number of different ways to chose
a symplectic basis, that we shall order as $\{e_1,\ldots,\e_n; f_1,\ldots, f_n \}$,
of the vector
space $\F_q^{2n}$, i.e.,  the bilinear form in this basis is
$(e_i,f_i)=1=-(f_i,e_i)$ and $(e_i,f_j)=0=(f_j,e_i)$ for $i\neq j$.

\begin{enumerate}
\item{Choose $e_1$: Any non-zero vector in $\F_q^{2n}$ could be chosen as $e_1$. The number of
	vectors in $\F_q^{2n}$ is $q^{2n}$ 	 so that the number of non-zero vectors is
	 $\# e_1 = q^{2n}-1$.  }
\item{Choose $f_1$: The vector $f_1$ might be chosen among the vectors which are not
	orthogonal to $e_1$. The dimension of the subspace $\{ v \in \F_q^{2n} | (v,e_1)=0 \}$ is $2n-1$
(at this point we could think of $e_1$ as being an element of an orthonormal basis of
$\F_q^{2n}$) so that the number of the orthogonal vectors is $q^{2n-1}$.  }
\item{The number of vectors which are not orthogonal to $e_1$ is thus $q^{2n} -q^{2n-1} =(q-1)q^{2n-1}$
and all they come in scalar multiples such as, $f_1, \ 2 f_1, \ldots, \ (q-1)f_1$, so that the number of
different choices for $f_1$ is $\# f_1=q^{2n-1}$. }
\item{Therefore the number of different choices of $e_1, f_1$ is
	\[
	\# \{e_1, f_1\}=(q^{2n}-1)q^{2n-1}  .
	\]
 }
\item{The rest of the space, which is spanned by the  basis vectors
	$\{e_2,\ldots,\e_n; f_2,\ldots, f_n \}$
is the vector space $\F_q^{2n-2}$ of dimension $2(n-1)$ so we can
 apply again the above result to obtain
 \[
	\# \{e_2, f_2\}=(q^{2(n-1)}-1)q^{2(n-1)-1}
	\]
(after $\{e_1,f_1\}$ have been chosen).
}
\item{Induction: continuing in this way we finally obtain
\beq \label{Sp-q}
 |Sp_{2n}(q)| = \prod_{j=n \downarrow}^1  (q^{2j}-1)q^{2j-1} = q^{n^2}  \prod_{j=1}^n  (q^{2j}-1)    .
\eeq }
\end{enumerate}
Applying Eq.~(\ref{Sp-q}) for $q=2$ we obtain Eq.~(\ref{Sp-2}).
\section{Ising braiding gates as symplectic transformations}
\label{app:B}
We will determine the symplectic matrices corresponding to the Ising model exchange matrices via relation (\ref{isomorphism}). To this end we first phrase the Ising model quantum computer abstractly in three steps:
\begin{itemize}
\item Starting point is the representation of the Clifford algebra in terms of the $\gamma_j^{(n+1)}$. According to (\ref{R}) we define operators $R_{j}$ which act on $\H^{\otimes n+1}$ where $\H=\C^2$. 
\item The subspace of $\H^{\otimes n+1}$ of positive parity, i.e. the linear span of the set
\[
\{|x_1 \dots x_{n+1}\ra\,:\,x_1\oplus\dots\oplus x_{n+1}=0\}\, ,
\]
will be denoted by $\H^{\otimes n+1}_+$. We will identify $\H^{\otimes n+1}_+$ with $\H^{\otimes n}$ via the correspondence
\[
|x_1 \dots x_n\,z\ra \leftrightarrow |x_1 \dots x_n\ra
\]
where $z$ is chosen such that the left hand side is an element of $\H^{\otimes n+1}_+$, i.e. $z=x_1\oplus\dots\oplus x_n$.
\item Finally, the action of the $R_{j}^+$ on $|x_1 \dots x_n\ra\in\H^{\otimes n}$ is obtained by applying $R_{j}$ to $|x_1 \dots x_n\,z\ra$ and calculating the corresponding vector in $\H^{\otimes n}$. Note that this step is only well-defined, because $R_{j}$ is reducible with respect to the decomposition $\H^{\otimes n+1} =\H^{\otimes n+1}_+ \oplus \H^{\otimes n+1}_-$.
\end{itemize}
The last two steps describe the action of the projection $P_{\pm}^{(n)}$ in section \ref{sec:exchange}.

Now, instead of focusing on the correspondence between basis vectors, we will concentrate on the correspondence between the Pauli groups associated with $\H^{\otimes n}$ and $\H^{\otimes n+1}$. The operators
\beq
\label{NewPauli1}
 \tau_j^{(k)} \equiv \I_{k-1} \otimes \sigma_j \otimes \I_{n-k}\otimes {\sigma_1}\upharpoonright_+
\eeq
if $j\neq 3$ and
\beq
\label{NewPauli2}
\tau_3^{(k)} \equiv \I_{k-1} \otimes \sigma_3 \otimes \I_{n-k}\otimes {\I}\upharpoonright_+\,,
\eeq
where we denote by $\upharpoonright_+$ the restriction to $\H^{\otimes n+1}_+$, are essentially the Pauli operators on $\H^{\otimes n}$. This follows from the fact that these operators are also reducible with respect to the decomposition $\H^{\otimes n+1}=\H^{\otimes n+1}_+ \oplus \H^{\otimes n+1}_-$ which allows us to calculate their action on $\H^{\otimes n}$ just like we do for the $R_j^+$. 

Since both,  $R_{j}$ and $R_{j}^+$ are Clifford unitaries in their respective domains, it is easy to determine the action of $R_{l}^+$ on $\tau_j^{(k)}$ according to the following commutative diagram
\[
\begin{array}{ccc}
(R_{l}^+)^* \tau_j^{(k)}R_{l}^+ & \longrightarrow & \displaystyle \prod_{r=1}^n \tau_{\alpha_r}^{(r)}\vspace{10pt}\\
\big\downarrow & & \big\uparrow \vspace{10pt}\\
R_{l}^* \sigma_j^{(k)}\sigma_{\delta}^{(n+1)}R_{l}\upharpoonright_+ \hspace{10pt}& \longrightarrow &\hspace{10pt}\displaystyle \prod_{r=1}^n \sigma_{\beta_r}^{(r)}\upharpoonright_+
\end{array}
\]
where we have used the short hand notation
\[
\sigma_i^{(k)}\equiv \I_{k-1}\otimes \sigma_i \otimes \I_{n-k}\,.
\]
Note that the $\sigma_{\delta}^{(n+1)}$ in the lower left corner depends on the $\sigma_j^{(k)}$, we have $\delta =1$ if $j=1,2$ and $\delta=0$ if $j=0,3$.

The action of the operators $R_{j}$ on Pauli operators is equivalent\footnote{The operators $R_j$ and $P,\,R$ differ only by a global phase factor, which means that the corresponding elements in the projective Clifford group $\P\Cl_n$ coincide.} to the following Clifford unitaries
\[
R_{2j-1} \simeq P = \left[ \begin {array}{cc} 
1&0\\
0&i\end {array}
\right ]
\]
\[
R_{2j} \simeq R =\frac{1}{\sqrt{2}} \left[ \begin {array}{cccc} 
1&0&0&i\\
0&1&-i&0\\
0&-i&1&0\\
i&0&0&1\end {array}
	\right]\, .
\]
These Clifford gates map Pauli operators according to
\[
\sigma_1 \stackrel{P}{\longrightarrow} \sigma_2\quad ,\quad \sigma_2 \stackrel{P}{\longrightarrow} \sigma_1
\]
\beqa
\sigma_1 \otimes \I \stackrel{R}{\longrightarrow}  \sigma_3\otimes \sigma_2 & \quad , \quad & \sigma_2 \otimes \I \stackrel{R}{\longrightarrow}  \sigma_2\otimes \I \nonumber\\
\I \otimes \sigma_1 \stackrel{R}{\longrightarrow}  \sigma_2\otimes \sigma_3& \quad , \quad & \I \otimes \sigma_2 \stackrel{R}{\longrightarrow}\I\otimes\sigma_2 \quad .\nonumber
\eeqa
It is straightforward to determine the action of the $R_{j}^+$ for all $j=1,\dots , 2n-1$. Since the corresponding $R_{j}$ do not effect qubit $n+1$ we get non-trivial relations only for qubits $j$ and $j+1$
\beqa
R_{2j-1}^+\,:\,& \tau_1^{(j)}\longrightarrow \tau_2^{(j)} &\quad , \quad \tau_2^{(j)}\longrightarrow \tau_1^{(j)}\nonumber\\
R_{2j}^+\,:\,& \tau_1^{(j)}\longrightarrow \tau_3^{(j)}\tau_2^{(j+1)} & \quad , \quad \tau_2^{(j)}\longrightarrow \tau_2^{(j)}\nonumber\\
& \tau_1^{(j+1)}\longrightarrow \tau_2^{(j)}\tau_3^{(j+1)} &\quad , \quad \tau_2^{(j)}\longrightarrow \tau_2^{(j)}\,.\nonumber
\eeqa
The action of the last two operators can be calculated by utilizing
\[
{\sigma_3^{\otimes n+1}}\upharpoonright_+ \equiv \I\upharpoonright_+ \,.
\]
It follows that
\[
\tau_1^{(k)}\stackrel{R_{2n+1}^+}{\longrightarrow} \tau_2^{(k)}\prod_{r\not =k}\tau_3^{(r)}
\]
\[
\tau_2^{(k)}\stackrel{R_{2n+1}^+}{\longrightarrow} \tau_1^{(k)}\prod_{r\not= k}\tau_3^{(r)}\, ,
\]
which can be verified as follows
\beqa
(R_{2n+1}^+)^*\tau_1^{(k)}R_{2n+1}^+ & = & (R_{2n+1}^+)^*\sigma_1^{(k)}\sigma_1^{(n+1)} R_{2n+1}^+\upharpoonright_+ \nonumber \\
& = & \sigma_1^{(k)}(R_{2n+1}^+)^*\sigma_1^{(n+1)}R_{2n+1}^+\upharpoonright_+ \nonumber \\
& = & \sigma_1^{(k)} \sigma_2^{(n+1)}\upharpoonright_+ \nonumber \\
& = & \sigma_1^{(k)} \sigma_2^{(n+1)}\upharpoonright_+ \I \upharpoonright_+ \nonumber \\
& = & \sigma_1^{(k)} \sigma_2^{(n+1)}\, \sigma_3^{\otimes n+1}\upharpoonright_+ \nonumber\\
& = & \sigma_3^{\otimes k-1} \otimes \sigma_2 \otimes \sigma_3^{\otimes n-k}\otimes \sigma_1 \upharpoonright_+ \nonumber
\eeqa
and similarly for $\tau_2^{(k)}$. The relations for $R_{2n}^+$ can be calculated in the same way, but we will skip the proof and give only the result. If $k\leq n-1$ we have
\[
\tau_1^{(k)}\stackrel{R_{2n+1}^+}{\longrightarrow} \left(\prod_{r\not= k,n}\tau_3^{(r)}\right) \tau_2^{(k)}\tau_1^{(n)}
\]
\[
\tau_2^{(k)}\stackrel{R_{2n+1}^+}{\longrightarrow} -\left( \prod_{r\not= k,n}\tau_3^{(r)}\right)\tau_1^{(k)}\tau_1^{(n)}
\]
The case $k=n$ gives
\[
\tau_1^{(n)}\stackrel{R_{2n+1}^+}{\longrightarrow} \tau_1^{(n)}
\]
\[
\tau_1^{(k)}\stackrel{R_{2n+1}^+}{\longrightarrow} \prod_{r=1}^{n}\tau_3^{(r)}\,.
\]
From these equations we can read off the corresponding symplectic matrices
\[
S_{2i-1}=\left[ \begin {array}{c|c|c} \I_{2(i-1)}&0&0\\
	\hline 0& \begin{footnotesize}\begin{array}{cc}0&1\\ 1&0 \end{array}\end{footnotesize} &0\\
	\hline 0&0&\I_{2(n-i)}\end{array}
 \right]
\]
for $1\leq 2i-1\leq 2n-1$ and
\[
S_{2i}=\left[ \begin {array}{c|c|c} \I_{2(i-1)}&0&0\\
	\hline 0& \begin{footnotesize}\begin{array}{cccc}1&0&0&0\\ 1&1&1&0\\0&0&1&0\\1&0&1&1 \end{array}\end{footnotesize} &0\\
	\hline 0&0&\I_{2(n-i-1)}\end{array}
 \right]
\]
for $1\leq 2i \leq 2n-2$. The last two matrices read
\[
S_{2n}=\left[ \begin{footnotesize}\begin {array}{cc|cc|c|cc} 0&1&1&1& \dots &0&1\\
	1&0&1&1 & & 0&1\\
	\hline 1&1&0&1 & & 0&1\\
	1&1&1&0 & & 0&1\\
	\hline \vdots & &  &  & \ddots & \vdots &\vdots \\
	\hline 1&1&1 &1 &\dots & 1&1\\
	0&0&0&0 & & 0&1\\\end{array}\end{footnotesize}
 \right]
\]
\[
S_{2n+1}=\left[ \begin{footnotesize}\begin {array}{cc|cc|cc|c} 
	0&1&1&1& 1&1&\dots  \\
	1&0&1&1 & 1&1&\\
	\hline 
	1&1&0&1 & 1&1&\dots\\
	1&1&1&0 & 1&1&\\
	\hline
	1&1&1&1&0&1&\\
	1&1&1&1&1&0&\\
	\hline
	\vdots & &\vdots  &  &  & &\ddots \end{array}\end{footnotesize}
 \right]\,.
\]
\section{Faithfulness of the representation of the permutation group}
\label{app:faith}
In this appendix we will use the symplectic description of the braid generators given in Appendix \ref{app:B} to 
prove that the equivalence classes
(i.e., the cosets of the monodromy subgroup) of the elementary braid generators are all different and hence they 
form a faithful representation of the symmetric group.
According to section \ref{sec:braiding} we have the relation
\[
\mathrm{Image}(\B_{2n+2}) /\mathrm{Image}(\M_{2n+2}) \subset Sp_{2n}(2)
\]
and therefore the advantage of the symplectic description is that symplectic matrices already describe 
classes of braiding operators with respect to the equivalence relation in which two elements are equivalent if they differ by a monodromy transformation only and therefore it is natural to compare these matrices directly.

Recall that a representation of the permutation group $\S_{2n+2}$ is by definition generated by $2n+1$  matrices
\[
S_j, \quad 1\leq j\leq 2n+1
\]
satisfying the relations 
\beqa 
S_j S_{j+1}S_j &=& S_{j+1}S_j S_{j+1}, \nn
S_j S_{k}&=& S_k S_j, \quad |k-j| >1,  \nn 
(S_j)^2&=& \I \nonumber
\eeqa
and this representation is faithful if and only if each generator $S_j$ is independent from the others with respect to the group multiplication. We denote the elementary generators of the abstract group $\S_{2n+2}$ by $\pi_j$. It is understood that each $S_j$ represents $\pi_j$.

It is easy to see that the matrices $S_j$, that we derived in Appendix \ref{app:B},
indeed satisfy the above relations. 
To further analyze the representation given by the $S_j$ it is convenient to change the basis in the symplectic space according to
\[
(\underbrace{0,\dots , 0}_{2i},1,0,\underbrace{0,\dots ,0}_{2(n-i-1)}) \rightarrow (\underbrace{0,\dots , 0}_{2i},1,0,\underbrace{1,\dots ,1}_{2(n-i-1)})
\]
\[
(\underbrace{0,\dots , 0}_{2i},0,1,\underbrace{0,\dots ,0}_{2(n-i-1)}) \rightarrow (\underbrace{0,\dots , 0}_{2i},0,1,\underbrace{1,\dots ,1}_{2(n-i-1)})\,.
\]
To obtain the matrices $S_j$ in this new basis we have to multiply them from both sides with the 
(self-inverse over $\F_2$) matrix
\[
T=\left[ \begin{footnotesize}\begin {array}{cc|cc|c|cc} 1&0&0&0& \dots &0&0\\
	0&1&0&0 & & 0&0\\
	\hline 1&1&1&0 & & 0&0\\
	1&1&0&1 & & 0&0\\
	\hline \vdots & &  &  & \ddots & \vdots &\vdots \\
	\hline 1&1&1 &1 &\dots & 1&0\\
	1&1&1&1 & & 0&1\\\end{array}\end{footnotesize} \right]
\]
which changes between both bases. The new symplectic matrices $\tilde S_j = T\cdot S_j\cdot T$ are 
\[
\tilde S_{2i-1}=\left[ \begin {array}{c|c|c} \I_{2(i-1)}&0&0 \\
	\hline 0& \begin{footnotesize}\begin{array}{cc}0&1\\ 1&0 \end{array}\end{footnotesize} & 0\\
	\hline 0&0&\I_{2(n-i)}\end{array}
 \right]
\]
for $1\leq 2i-1 \leq 2n-1$,
\[
\tilde S_{2i}=\left[ \begin {array}{c|c|c} \I_{2i-1}&0&0\\
	\hline 0&  \begin{footnotesize}\begin{array}{cc} 0&1\\ 1&0 \end{array}\end{footnotesize} &0\\
	\hline 0&0&\I_{2(n-i)-1}\end{array}
 \right]
\]
for $1\leq 2i \leq 2n-2$ and the last two are
\[
\tilde S_{2n}=\left[ \begin{footnotesize}\begin {array}{cccccc|c} 1&0&0&0& \dots &0&1\\
	0&1&0&0 & & 0&1\\
	0&0&1&0 & & 0&1\\
	0&0&0&1 & & 0&1\\
	\vdots & &  &  & \ddots & \vdots &\vdots \\
	0&0&0 &0 &\dots & 1&1\\
	\hline 0&0&0&0 & & 0&1\\\end{array}\end{footnotesize}
 \right]
\]
\[
\tilde S_{2n+1}=\left[ \begin{footnotesize}\begin {array}{cc|cc|cc|c} 
	0&1&1&1& 1&1&\dots  \\
	1&0&1&1 & 1&1&\\
	\hline 
	1&1&0&1 & 1&1&\dots\\
	1&1&1&0 & 1&1&\\
	\hline
	1&1&1&1&0&1&\\
	1&1&1&1&1&0&\\
	\hline
	\vdots & &\vdots  &  &  & &\ddots \end{array}\end{footnotesize}
 \right]\,.
\]
It is now obvious that the $\tilde S_j$ for $j=1,\dots , 2n-1$ represent elementary transpositions. Moreover, it is clear that $\tilde S_{2n}$ can never be constructed solely from the $\tilde S_j$ for $j=1,\dots,2n-1$ because of its non-trivial last column. Therefore the operators $\tilde S_{1}, \dots ,\tilde S_{2n}$ generate a faithful representation of $\S_{2n+1}$. 

If $n=1$ the two $2\times 2$ matrices $\tilde S_3$ and $\tilde S_1$ coincide, hence we get a faithful representation of $\S_3$. However,  for $n\geq 2$ a simple argument shows that $\tilde S_{2n+1}$ is independent from all
other $\tilde S_j$. Indeed, suppose there is a way of writing $\tilde S_{2n+1}$ as a product of the $\tilde S_j$ with $j\leq 2n$
\[
\tilde S_{2n+1}= \tilde S_{\alpha_1}\cdot \dots \cdot \tilde S_{\alpha_r}\,,\quad 1\leq \alpha_i \leq 2n\,.
\]
The right hand side corresponds to a permutation $\pi \in \S_{2n+2}$ whose expression in elementary group generators $\pi_j$ reads
\[
\pi\equiv \pi_{\alpha_1}\cdot\dots\cdot\pi_{\alpha_r}\,.
\]
Alternatively, the element $\pi$ can always be written as
\[
\pi \equiv \left( \prod_s \pi_{\alpha_s}\right)\pi_{2n}\left(\prod_t \pi_{\beta_t} \right)\, ,\quad \alpha_s,\,\beta_t \leq 2n-1\,.
\]
Therefore we also have the relation
\[
\tilde \S_{2n+1} \equiv \left( \prod_s \tilde S_{\alpha_s}\right)\tilde S_{2n}\left(\prod_t \tilde S_{\beta_t} \right)\,.
\]
Now observe that multiplying $\tilde S_{2n}$ with $\tilde S_j$'s with $j\leq 2n-1$ simply exchanges rows and columns in $\tilde S_{2n}$. The number of $1$'s in $\tilde S_{2n}$ is left unchanged by exchanging rows or columns. Therefore, it is obvious that $\tilde S_{2n+1}$ cannot be constructed like that from $\tilde S_{2n}$, because the number of $1$'s of both matrices differs. Hence, we have shown that the representation of the permutation group $\S_{2n+2}$ is faithful for all $n\geq 2$.

\end{appendix}
\bibliography{Z_k,my,FQHE,TQC,QI}
\end{document}